\pdfoutput=1

\documentclass[a4paper,fleqn,usenatbib]{article}

\usepackage{amsmath}	% Advanced maths commands

\usepackage{amssymb}	% Extra maths symbols
\usepackage{txfonts}

\usepackage{hyperref}
\usepackage[mathcal]{euscript}

\usepackage[T1]{fontenc}
\usepackage{ae,aecompl}

\usepackage[a4paper,margin=2.5cm]{geometry}
\usepackage{cite}

\usepackage{graphicx}	% Including figure files
\usepackage{subfig}
\usepackage{color}

\usepackage[utf8]{inputenc}
\usepackage{authblk}

\DeclareMathAlphabet{\mathpzc}{OT1}{pzc}{m}{it}

\begin{document}

\title{Periodic equatorial orbits in a black bounce scenario}
\author[1]{Anderson Bragado\thanks{E-mail: andersonbragado@fisica.ufc.br}}

\affil[1]{Universidade Federal do Cear\'a (UFC), Departamento de F\'isica,
Campus do Pici, Fortaleza - CE, C.P. 6030, 60455-760 - Brazil}

\author[2,1]{Gonzalo J. Olmo\thanks{E-mail: gonzalo.olmo@uv.es}}
\affil[2]{Departamento de F\'{i}sica Te\'{o}rica and IFIC, Centro Mixto Universidad de Valencia - CSIC. Universidad
de Valencia, Burjassot-46100, Valencia, Spain.}

\renewcommand\Authands{ and }

\label{firstpage}
\maketitle

\begin{abstract}
We study equatorial closed orbits in a popular black bounce model to see if the internal structure of these objects could lead to peculiar observable features. Paralleling the analysis of the Schwarzschild and Kerr metrics, we show that in the black bounce case each orbit can also be associated with a triplet of integers which can then be used to construct a rational number characterizing each periodic orbit. When the black bounce solution represents a traversable wormhole, we show that the previous classification scheme is still applicable with minor adaptations. We confirm in this way that this established framework enables a complete description of the equatorial dynamics across a spectrum of cases, from regular black holes to wormholes.  Varying the black bounce parameter $r_{min}$, we compare the trajectories in the Simpson-Visser model with those in the Schwarzschild metric (and the rotating case with Kerr). We find that in some cases even small increments in $r_{min}$ can lead to significant changes in the orbits.
\end{abstract}


%%%%%%%%%%%%%%%%%%%%%%%%%%%%%%%%%%%%%%%%%%%%%%%%%%

%%%%%%%%%%%%%%%%% BODY OF PAPER %%%%%%%%%%%%%%%%%%

\section{Introduction}

\label{sec:intro}

Recent observational results, such as images of the supermassive black hole candidates M87* and SgrA* captured by the Event Horizon Telescope (EHT) \cite{collaboration2019first, akiyama2022first}, findings by the GRAVITY collaboration \cite{abuter2018detection, abuter2020detection}, and the detection of gravitational waves by the Laser Interferometer Gravitational-Wave Observatory (LIGO) and Virgo collaborations \cite{abbott2016observation, abbott2017gw170817}, along with the expectations for the upcoming LISA mission \cite{barausse2020prospects}, have intensified interest in alternative compact objects that can mimic black holes. Examples of such alternatives include boson stars \cite{guzman2009spherical, herdeiro2021imitation}, gravastars \cite{Ray:2020yyk}, regular black holes \cite{bardeen1968non, roman1983stellar, hayward2006formation, frolov2014information, lan2023regular}, and wormholes \cite{damour2007wormholes, bueno2018echoes}.

At the same time, from another point of view, the inevitability of singularities in black hole solutions \cite{penrose1965gravitational, hawking1970singularities} is usually regarded as a fundamental problem in General Relativity. In this sense, the possibility of the existence of non-singular models that have a behavior similar to black holes (except at the core, of course) is, by itself, interesting \cite{carballo2018phenomenological, carballo2020opening, carballo2020geodesically}. These possible deviations from the classical predictions would usually be justified as the consequences of some yet unknown quantum gravitational effect or other kind of mechanism. Hence, the study of such geometries and their phenomenological consequences may ultimately provide a more complete model, pushing the boundaries of physics.

Here we consider an alternative compact object model proposed by Simpson and Visser \cite{simpson2019black}, known as {\it black bounce}, to study the properties of periodic orbits of material particles in the static, spherically symmetric case and also in its rotating version. The Simpson-Visser metric is a minimal modification of the Schwarzschild solution and it neatly interpolates between it, regular black holes, and traversable wormholes, depending on the value of a parameter that we shall denote as $r_{min}$. In the same way, its axisymmetric counterpart can be described with a similar modification of the Kerr metric, generated by employing the Newman-Janis procedure \cite{newman1965note,newman1965metric,mazza2021novel}, and can also represent different types of spacetimes.

In this context, some particular aspects of geodesics have already been studied in the Simpson-Visser model \cite{stuchlik2021epicyclic, de2024orbits}, and its rotating version has also been investigated \cite{mazza2021novel,islam2021strong,nosirov2023particle, jiang2021testing}. However,  a general classification of trajectories in these models is still lacking, primarily because of the complexity of the orbits in such backgrounds. The purpose of this paper is to make some progress filling this gap for time-like orbits in the Visser-Simpson model and use it to describe equatorial orbits in its rotating generalization.

To proceed, we will use a taxonomy for periodic orbits based on zooms, whirls, and vertices (purely topological features) proposed by Levin and Perez-Giz \cite{levin2008} for orbits in the Schwarzchild and Kerr metrics. As we will see, this approach sets a connection between closed orbits and rationals that will  completely describe the equatorial dynamics. In this context, we will show that both rotating and non-rotating Simpson-Visser metrics preserve the essential features needed for this classification and study how $r_{min}$ affects the behavior of these geodesics across different regimes, exploring the analogies existent between black bounce models and the Schwarzschild and Kerr solutions.

Throughout this paper we use geometrized units ($G = c = 1$) and, for most part of it, we set the black hole mass $M$ to one. Also, the reader should know that every orbit is specified by the energy $E$ and angular momentum $L$ of a test particle as measured by an observer at infinity. With our choice of units, all parameters can be treated as dimensionless.

\section{Classification of orbits}
\label{sec:classification}

The types of trajectories that massive particles can have in Newtonian gravity are well known: one can find elliptic orbits and hyperbolic or parabolic trajectories. However, in the context of General Relativity, complications arise and one needs a more robust approach in order to classify the different outcomes. In this sense, the possible orbits in strong-field regimes, where relativistic effects are relevant, are substantially different from the Newtonian ones, presenting, for instance, features we will refer to as leaves and whirls. In fact, assuming spherical symmetry, we will next see that a relatively simple classification scheme is possible.

With this in mind, following the approach outlined by Levin and Perez-Giz \cite{levin2008} we will next summarize how each closed orbit around a black hole can be associated with three integers: $z$, $w$ and $v$ (which rely solely on the topological features of the orbits). Using these numbers, one can construct a rational number that serves as a label for different types of closed orbits. 

\subsection{The integers $(z, w, v)$}

\begin{figure}[h]
\centering
 \includegraphics[width=0.5\columnwidth]{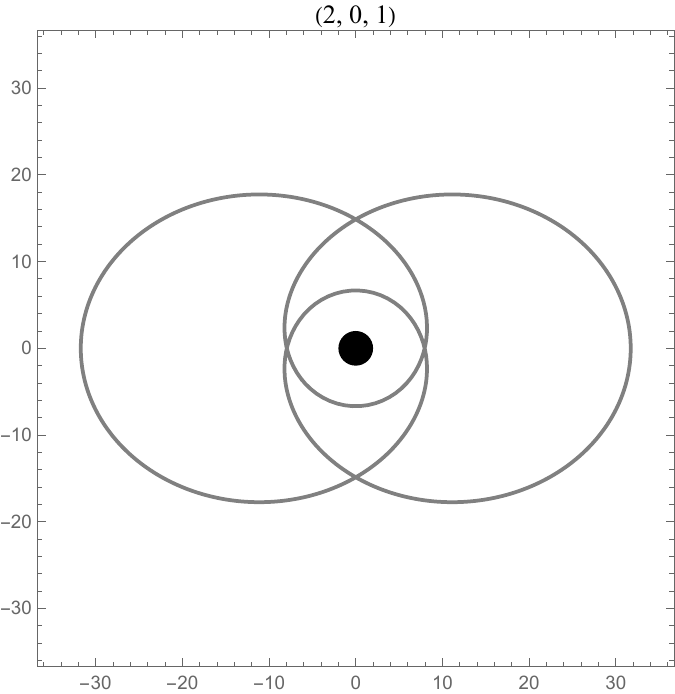}
 \caption{Periodic orbit in Schwarzschild black hole (the black circle represents the event horizon) with the triplet $(z = 2, w = 0, v = 1)$. Specifically, this orbit has $L = 4$ and $E = 0.975624$.}
 \label{fig:2-leaf}
\end{figure}

The first integer, $z$, is the number of leaves the orbit traces before closing. We call this number $z$ for zoom (since the orbits present this characteristic of zooming in and out, that actually defines what a leaf is). Figure \ref{fig:2-leaf} illustrates an orbit with $z = 2$. In the Newtonian context, we would simply classify an ellipse with $z = 1$, but as the relativistic effects start to be more relevant, these orbits will slowly start to precess (as in the case of Mercury's trajectory) and the orbits would be labeled with a very large number of leaves $z$, because a precessing ellipse can be regarded as a many-leaves orbit. As we enter in strong-field regimes, we introduce not only small perturbations (causing precessions) and the pattern of leaves can be more easily discerned.

\begin{figure}[h]
\centering
 \includegraphics[width=0.5\columnwidth]{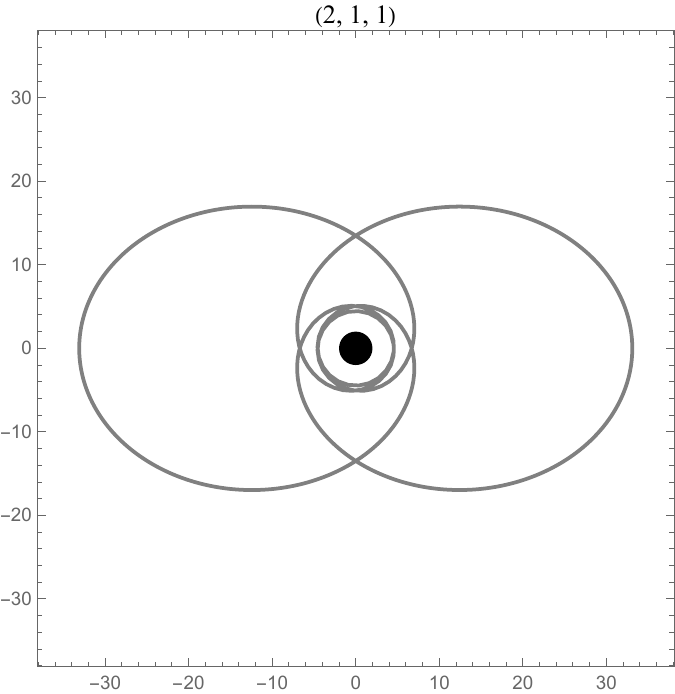}
 \caption{Periodic orbit in Schwarzschild black hole with the triplet $(z = 2, w = 1, v = 1)$. Specifically, this orbit has $L = 3.8$ and $E = 0.975685$.}
 \label{fig:2-leaf_w=1}
\end{figure}

However, this is not the only kind of $z = 2$ orbit. Actually, the particle can whirl around the center before zooming out to the apastron. Indeed, there are, in general, orbits that whirl around the center an angle $2\pi w$ before zooming out again. Thus, we also need the number of whirls, $w$, to describe the orbit. Figure \ref{fig:2-leaf_w=1} provides an example of a 2-leaf orbit with $w = 1$. We point out that this feature is not observed in the context of weak field regimes, thus no analogy can be drawn with Newtonian orbits, indicating the intrinsically different families of trajectories existent in General Relativity.

\begin{figure}[h]
\centering
 \includegraphics[width=1\columnwidth]{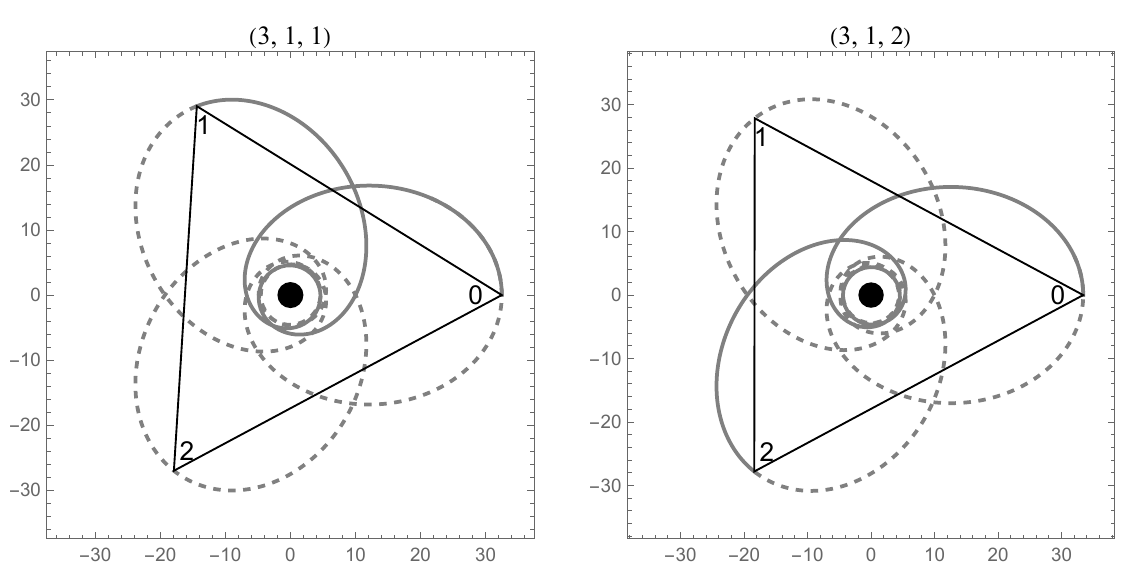}
 \caption{Periodic orbits in Schwarzschild black hole. Left: orbit with the triplet $(z = 3, w = 1, v = 1)$, with parameters $L = 3.8$ and $E = 0.975342$. Right: orbit with the triplet $(z = 3, w = 1, v = 2)$, with parameters $L = 3.8$ and $E = 0.975858$.}
 \label{fig:3-leaf_dif_v}
\end{figure}

Nevertheless, we still need another integer number. Notice that we can see the apastra of a $z > 2$ orbit as the vertices of a regular polygon. We label these vertices by the integer $v$, considering the initial apastron as $v = 0$ and increasing it in the same rotational sense as the orbit (counterclockwise for prograde orbits and clockwise for retrograde). Considering this, a given $z > 2$ orbit can go from the initial apastron right to the next (in which case $v = 1$) or, for instance, skip it and go to the next neighbor (with $v = 2$). More generally, the particle can skip any number of vertices $v$ less than $z$ for a given closed orbit. Figure \ref{fig:3-leaf_dif_v} illustrates this for $z = 3$ and $w = 1$ orbits. With these examples, we established the triplet $(z, w, v)$ that specifies each orbit.

Although the vertices of 2-leaf orbits do not trace a polygon, we can address them with $v = 1$. In addition, we label 1-leaf orbits with $v = 0$ (the only case where $v$ assumes this value). We point out that there may be different pairs of $(z,v)$ that would describe the same orbit for a given $w$. For example, the $(4,1,2)$ orbit is equivalent to the $(2,1,1)$ orbit. To remove this degeneracy, we impose $z$ and $v$ to be relatively prime. Therefore, with these restrictions, the triplet $(z,w,v)$ determines uniquely the topological aspects of a closed orbit, where the allowed $v$ are:
\begin{eqnarray}
\begin{aligned}
1 \leq v \leq z - 1, & \quad \text{if } z > 1, \text{ with $z$ and $v$ relatively prime}, \\
v = 0, & \quad \text{if } z = 1.
\end{aligned}
\label{v_lim}
\end{eqnarray}

\subsection{Relation with rational numbers}
With the triplet $(z,w,v)$ we can naturally see the relation between periodic orbits and rationals. First note that any rational number can be written as:
\begin{equation}
q = s + \frac{m}{n},
\end{equation}
where $s \geq 0$ is the integer part and $m/n$ is the fractional part,
with $m$ and $n$ relatively prime integers satisfying
\begin{equation}
1\le m\le n-1.  
\end{equation}
These are exactly the same relations as those satisfied by $z$, $w$ and $v$. Thus, every periodic orbit is associated to the rational number:
\begin{equation}
  q \equiv w + \frac{v}{z}
  \label{qdef}
\end{equation}
This rational number actually has a physical meaning: it is related to the accumulated azimuth between successive apastra $\Delta \varphi_r$. In this sense, as can be easily seen, we can write:
\begin{equation}
\Delta \varphi_r=2\pi \left(1+w+\frac{v}{z}\right )=\frac{\Delta \varphi}{z},
\label{dphir}
\end{equation}
where the total accumulated angle $\Delta \varphi$ in one full orbital
period is $z\Delta \varphi_r$. As we shall show, this relation provides a basis for finding the initial conditions that determine orbits with a given associated rational number $q$.

Nevertheless, there is another well-known way of relating periodic orbits and rational numbers \cite{poincare1892, Landau1976Mechanics}. An eccentric equatorial orbit has a radial frequency
\begin{equation}
\omega_r=\frac{2\pi}{T_r},
\end{equation} 
where $T_r$ is the (coordinate) time elapsed during one radial
cycle (not the total period of an orbit), and an angular frequency
\begin{equation}
\omega_\varphi=\frac{1}{T_r} \int_0^{T_r}\frac{d\varphi}{dt} dt =
\frac{\Delta\varphi_r}{T_r}  \quad\quad
\end{equation}
Of course, for the movement to repeat itself we need the ratio $\omega_\varphi / \omega_r$ to be rational. In fact, we can see that, in this case:
\begin{equation}
  \frac{\omega_\varphi}{\omega_r} = \frac{\Delta\varphi_r}{2\pi} =
  1 + w +\frac{v}{z} = 1 + q
\end{equation}
Thus, also establishing a relation between $q$ and the orbital frequencies $\omega_r$ and $\omega_\varphi$. \\

Before concluding this section, it is important to note that even non-periodic orbits can be described by the above scheme, because we can approximate arbitrarily well an irrational number by a rational. Also, in principle, circular orbits do not fit in this description because they have no radial frequency to relate with $\omega_\varphi$.
In spite of this, we can associate a rational number to them by first noticing that the zero eccentricity limit of $\omega_r$ for stable circular orbits is not zero, but rather the frequency of radial oscillations for small perturbations \cite{levin2008}. 

\section{General behavior of $q$ in the Schwarzschild and Kerr cases}

We will now guide the reader on the qualitative relation that exists between different types of orbits and the range of values that the rational $q$ can take. For this purpose, we will have to explore the critical points of the effective potential of time-like geodesics, which will provide us with crucial information about the different types of orbits that can be found.\\

\subsection{Range of $q$ in the Schwarzschild background}

%Furthermore, unstable circular orbits also exist in the Schwarzschild and Kerr metrics. \cite{hobson2006general, wald1984general}. When those orbits are bound (meaning their energy is less than one), there will always be a homoclinic orbit associated to them. Such an orbit can be seen as one that, starting from a finite apastron, asymptotically approaches the corresponding unstable circular orbit, executing an infinite number of whirls and thereby having an associated rational number $q = \infty$. Those orbits have the same angular momentum $L$ and energy $E$ as the unstable circular orbit they correspond to. We explain these aspects in more detail next. \\

%In this case, as we shall see, the stable circular orbit will determine the minimum value $q$ could take, and if the unstable circular orbit is unbounded ($E > 1$) there will be also a superior limit (and otherwise $q$ can take arbitrarily large values). To show it, let us consider the effective potential for time-like geodesics in the Schwarzschild metric. 

Let us consider the effective potential for time-like geodesics in the Schwarzschild metric.
Considering the black hole mass $M$ equal to one, the motion can be described by \cite{wald1984general}:
\begin{equation}\label{eq:GeoSch}
\frac{1}{2}\dot r^2+V_{\rm eff}=\varepsilon_{\rm eff} \ ,
\end{equation}
where
\begin{equation}
  \varepsilon_{\rm eff} =\frac{E^2}{2} \, , \quad V_{\rm eff} = \frac{1}{2} -\frac{1}{r}+\frac{L^2}{2r^2}-\frac{L^2}{r^3}.
\label{veff_schw}
\end{equation}
Like in any problem of a particle in a potential, the range of $r$ where the particle can be found depends on the relation between $\varepsilon_{eff}=E^2/2$ and $V_{eff}$. This will determine the existence of bounded/unbounded orbits and critical cases. \\

\begin{figure}[!h]
\centering
 \includegraphics[width=0.8\columnwidth]{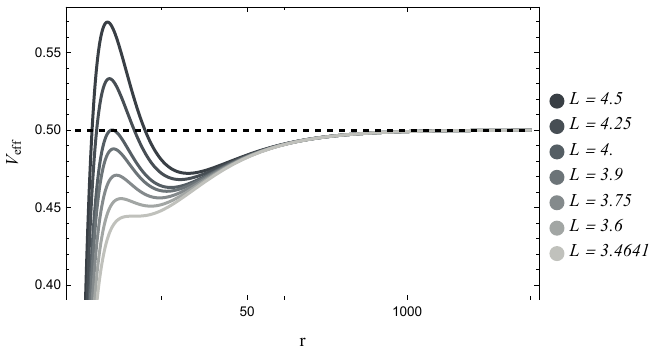}
 \caption{The effective radial potential $V_{\rm eff}(r)$ (plotted in a $\log{r}$ scale) for different values of angular momentum $L$. The horizontal line at $V_{\rm eff} = 0.5$ corresponds to the bound energy of $E = 1$. Note that $L_{IBCO} = 4$ and the lower curve have angular momentum $L_{ISCO} = \sqrt{12}$.}
 \label{fig:Veff}
\end{figure}

Since $V_{\rm eff}$ does not depend on the energy $E$, we can analyze the family of possible orbits just by looking at the plot of $V_{\rm eff}$ for each value of the angular momentum $L$. Consider Fig. \ref{fig:Veff}, where we show $V_{\rm eff}$ from large values of $L$ to low ones. The critical points in the potential, which are given by $dV_{\rm eff}/dr = 0$, determine the region where circular orbits are possible.  Solving this condition, we find:
\begin{equation}
    r_c^\pm= \frac{1}{2} \left(L^2 \pm L \sqrt{L^2-12}\right)
\label{circular_radii_schw}
\end{equation}

First of all, note that the critical points $r_c^\pm$ only exist for $L \geq \sqrt{12}$. A glance at Fig. \ref{fig:Veff} confirms that $r_c^+$ represents a local minimum, while $r_c^-$ is a local maximum. Naturally, {\bf stable circular orbits} occur at local minima, while {\bf unstable circular orbits} are found at local maxima. Additionally, when $L=\sqrt{12}$, the two critical points coincide ($r_c^+=r_c^-$) and besides having $dV_{\rm eff}/dr = 0$, one also has that $d^2V_{\rm eff}/dr^2 = 0$, which is the condition for {\bf marginal stability} \cite{hobson2006general}.
Because this is the stable circular orbit with smallest radius, we call it the {\bf Innermost Stable Circular Orbit} (ISCO), with angular momentum $L_{ISCO} = \sqrt{12}$ and associated radius $r_{ISCO} = 6$. The lowest curve in Fig. \ref{fig:Veff} represents the potential that contains this orbit. \\

Another important orbit to which we will refer to is the {\bf Innermost Bound Circular Orbit} (IBCO), which corresponds to a specific unstable circular orbit that we introduce now. As the angular momentum increases from $L=\sqrt{12}$, we see that the radius of the unstable orbit (at the maximum of the potential) decreases, and the value of the local maximum of the effective potential increases (see Fig. \ref{fig:Veff}). %This behavior can be easily demonstrated using equations \ref{veff_schw} and \ref{circular_radii_schw}.
Because of this, there exists a specific value of the angular momentum, $L=L_{IBCO}$, for which the effective potential at the local maximum coincides with its asymptotic value $\lim_{r \to \infty} V_{\rm eff}=1/2$.  This orbit is what we call the IBCO. For any angular momentum greater than $L_{IBCO}$, the corresponding unstable circular orbit becomes unbounded because any particle capable of reaching $r_c^-$ will have $E>1$ and will no longer be trapped within the potential walls. The importance of the IBCO in our context lies in the fact that bound maxima in the effective potential, those with $L<L_{IBCO}$ (for which $V_{eff}(r_c^-)<1/2$), can be associated with {\bf homoclinic orbits}. 

Homoclinic orbits are those that start at a finite apastron with the right energy to asymptotically approach the corresponding unstable circular orbit, executing an infinite number of whirls and thereby having an associated rational number $q \to  \infty$. Those orbits have the same angular momentum $L$ and energy $E$ as the unstable circular orbit they tend to. This means that closed orbits that asymptotically approach the homoclinic orbit can have 
arbitrarily large associated rational numbers. In contrast, for any $L>L_{IBCO}$, the associated rational number $q$ will have an upper bound. We thus see that the possible  rational numbers $q$ will be in the range 
\begin{center}
\begin{tabular}{lc}
$q_c \leq q \leq \infty$ &  \text{if $L_{ISCO} < L < L_{IBCO}$}, \\
$q_c \leq q \leq q_{max} $ & \text{if $L > L_{IBCO}$},
\end{tabular} 
\end{center}
\noindent where $q_c$ is the rational associated with the stable circular orbit and $q_{max}$ the rational associated with the bound orbit with maximum energy.\\

For the Schwarzschild geometry, the parameters that specify the IBCO are easily calculated as $L_{IBCO} = 4$ and $r_{IBCO} = 4$ (recall that we are setting $M=1$). In Figure \ref{fig:Veff}, the evolution of the effective potential is shown, from $L_{ISCO} = \sqrt{12}$ to $L_{IBCO} = 4$ and beyond, including values of $L$ that lead to unbounded, unstable circular orbits.

%For each value of $L\geq \sqrt{12}$, the stable circular orbit always corresponds to the lowest possible value for $E$. By contrast, when $L>4$, the unstable circular orbit (tip of the potential) will necessarily have $E > 1$. As a result, these orbits cannot approximate homoclinic ones and, consequently, they have a bound value for $q$. However, for a certain value of angular momentum, which we call $L_{IBCO}$ (the angular momentum for the Innermost Bound Circular Orbit), the maximum of the effective potential begins to fall below the value of  $V_{\rm eff}$ at infinity. This allows the orbits to approximate homoclinic ones, and as a result, there will be no upper limit for $q$. This behavior continues for $L < L_{IBCO}$ until $L = L_{ISCO}$ (the angular momentum for the Innermost Stable Circular Orbit), because for $L < L_{ISCO}$ closed orbits do not exist anymore. 

\subsection{Range of $q$ in Kerr and comments on the numerical approach}

Although Kerr orbits do not admit a simple one-dimensional effective potential, their description is qualitatively analogous to that of the Schwarzschild case. We still have an ISCO and an IBCO, the unstable circular orbit (when it exists) is always more energetic than the stable one, etc. \cite{levin2008}. Because of those features, the same methodology as we used to specify and study the Schwarzschild orbits is still valid.\\

\begin{figure}[!h]
\centering
 \includegraphics[width=0.8\columnwidth]{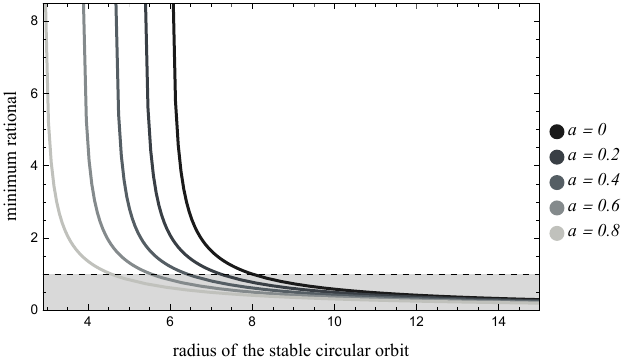}
 \caption{Lower bound of the associated rational $q$ versus the radius of the (co-rotating) stable circular orbit in the Kerr metric, shown for various values of the spin parameter $a$. Above the horizontal line at $y = 1$, all orbits must exhibit whirls, a feature exclusive to strong-field gravity near the black hole. As the ISCO is approached, the minimum diverges due to the merger of the unstable and stable circular orbits. As $a$ increases, the ISCO radius decreases for co-rotating orbits (and increases for counter-rotating orbits). 
 \label{fig:qmin}}
\end{figure}

Additionally, since the reasoning involving unstable circular orbits and homoclinic orbits determines if $q$ will have an upper bound, for completeness, it should be noted that the lower bound will always be determined by the value of $q$ corresponding to the stable circular orbit. In this sense, for large values of the radius of stable circular orbits (in the weak-field regime), $q$ tends to zero, reflecting the fact that orbits will have no whirls and start to resemble simple elliptical precession. As we approach the ISCO and the extrema of the potential begin to merge, the value of $q$ diverges. Fig. \ref{fig:qmin} shows this relation in the Kerr metric for different values of the spin parameter $a$.\\

Another essential feature is that $q$ increases monotonically with energy (since $\Delta \varphi_r$ does so) and decreases monotonically with angular momentum \cite{levin2008, levin2009energy}. In fact, this is the relation we will use  to determine the initial conditions for an orbit with a given associated rational in the next sections. Shortly, if, for instance, we maintain $L$ constant, denoting $r_a$ as the radius of the apastron and $r_p$ of the periastron, the process is to compute numerically the integral
\begin{equation}
\Delta \varphi_r=2\int_{t(r_p)}^{t(r_a)}\frac{d \varphi}{dt} dt = 2\int_{r_p}^{r_a}
\frac{d\varphi}{dr} dr
\label{eq:delta_phi}
\end{equation}
and use the bisection method to find the energy that returns the required rational $\Delta \varphi_r$ [see Eq.(\ref{dphir})]. In this process, we take advantage of the behavior of $q$ with respect to $E$, noting that the smallest $E$ corresponds to the stable circular orbit and the largest $E$ to the energy of the unstable circular orbit or to $E=1$. Denoting the lower and upper values for the energy as $E_1$ and $E_2$, we use the mean $(E_1 + E_2)/2$ as a trial energy to evaluate the integral in Eq.\ref{eq:delta_phi}. If the resulting angle is larger/smaller than the desired value, we set this trial energy as the new upper/lower limit and repeat the process. This is done iteratively until the angle converges to the specified precision. Of course, an analogous process would be valid if $L$ is varied and $E$ is kept constant.\\

\begin{figure}[!h]
\centering
\includegraphics[width=1\columnwidth]{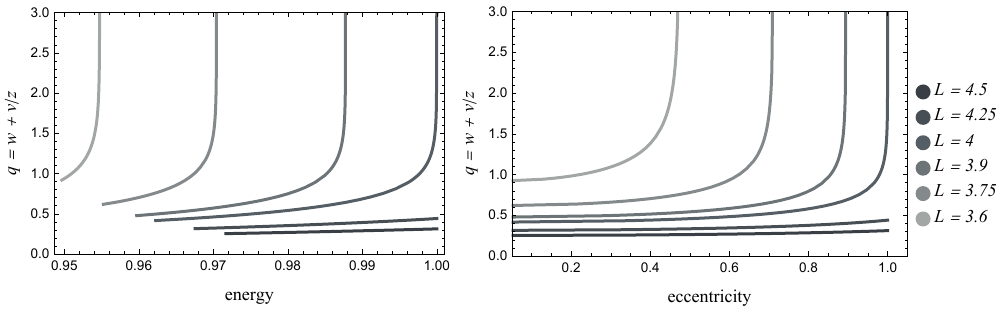}
 \caption{Curves of the associated rational vs energy and eccentricity in a Schwarzschild black hole. The different angular momentum are the same as Figure \ref{fig:Veff}, except for the $L = L_{ISCO}$.}
 \label{fig:qcurves}
\end{figure}

It is worth noting that $q$ also grows monotonically with the eccentricity $e$, defined as:
\begin{equation}
e = \frac{r_a - r_p}{r_a + r_p} \ ,
\end{equation}
where $r_a$ is the apastron and $r_p$ the periastron. This observation provides an alternative route to relating $q$ to observable quantities. % (instead of $E$ and $L$). 
Fig. \ref{fig:qcurves} compares the dependence of $q$ on the energy $E$ and the eccentricity $e$ for a Schwarzschild black hole, using the same values of angular momentum as in Fig. \ref{fig:Veff}. Note that, regardless of the representation, $q$ increases as $L$ decreases.
Note also that in the curves with $L > L_{IBCO}$ ($L = 4.25$ and $L = 4.5$), the value of $q$ is almost independent of $e$. This is because orbits with $L > L_{IBCO}$ cannot approach homoclinic ones. One can show that as $L$ grows beyond $L_{IBCO}$, the range of $q_{min}$ and $q_{max}$ becomes even smaller. As $L \to L_{IBCO}^-$, the allowed orbits execute many whirls (potentially infinite), leading to large values of $q$. In the IBCO case, orbits with large $q$ also have a large apastron radius, but as $L$ diminishes, the maximum drops and $q$ begins to grow for smaller values of $r_a$, divergingfor some $e < 1$ (or $E < 1$). For more examples of orbits in Schwarzschild and Kerr metrics see Figs. 11, 12, 14 and 15 in \cite{levin2008}.

\section{Orbits in the Simpson-Visser metric}
\label{sec:SV}
The metric proposed by Simpson and Visser \cite{simpson2019black} is given by:
\begin{equation}
    ds^{2}=-\left(1-\frac{2M}{\sqrt{r^{2}+r_{min}^{2}}}\right)dt^{2}+\frac{dr^{2}}{1-\frac{2M}{\sqrt{r^{2}+r_{min}^{2}}}}+\left(r^{2}+r_{min}^{2}\right)\left(d\theta^{2}+\sin^{2}\theta d\phi^{2}\right),
\label{eq:Simpson-Visser metric}
\end{equation}
where $M$ is the ADM mass and $r_{min}$ is a parameter that regularizes the central singularity. Note that for $r_{min} = 0$ this is simply the Schwarzschild solution. When  $r_{min}\neq 0$, it is possible to interpolate between regular black holes \cite{bardeen1968non, roman1983stellar, hayward2006formation, frolov2014information, lan2023regular} and traversable wormholes \cite{simpson2021traversable}. In particular, if $0 < r_{min} < 2M$ we have a regular black hole geometry with a one-way spacelike throat that lies within the event horizon, which is located at $r_H=\sqrt{4M^2 - r_{min}^2}$. If $r_{min} = 2M$, then we have a one-way wormhole geometry with an extremal null throat. For $r_{min} > 2M$ the geometry describes a traversable wormhole (in the sense of Morris-Thorne \cite{morris1988wormholes_am_phys, morris1988wormholes_phys_rev}). Throughout this section, we analyze the impact of  $r_{min}$ on the aspect of closed orbits in the region $r > 0$.\\

Let us first derive the equations governing the motion of massive test particles in the geometry defined above. 
Considering the vector tangent to the worldline of a massive particle parameterized by some $\lambda$ (that we choose to be the proper time), it follows that
\begin{equation}
    -1= g_{\mu \nu}\frac{dx^{\mu}}{d\lambda}\frac{dx^{\nu}}{d\lambda}=-g_{tt}\left(\frac{dt}{d\lambda}\right)^{2}+g_{rr}\left(\frac{dr}{d\lambda}\right)^{2}+\left(r^{2}+r_{min}^{2}\right)\left\lbrace\left(\frac{d\theta}{d\lambda}\right)^{2}+\sin^{2}\theta \left(\frac{d\phi}{d\lambda}\right)^{2}\right\rbrace \ .
\end{equation}
Because of the spherical symmetry, without loss of generality we can fix $\theta = \pi/2$, staying with the equatorial problem
\begin{equation}
    g_{\mu \nu}\frac{dx^{\mu}}{d\lambda}\frac{dx^{\nu}}{d\lambda}=-g_{tt}\left(\frac{dt}{d\lambda}\right)^{2}+g_{rr}\left(\frac{dr}{d\lambda}\right)^{2}+\left(r^{2}+r_{min}^{2}\right)\left(\frac{d\phi}{d\lambda}\right)^{2} = -1
    \label{geodesic}
\end{equation}
Now we can use the Killing symmetries of the line element \cite{wald1984general} to find  expressions for the conserved energy $E$ and conserved angular momentum $L$ (setting $M = 1$):
\begin{equation}
    \left(1-\frac{2}{\sqrt{r^{2}+r_{min}^{2}}}\right)\left(\frac{dt}{d\lambda}\right)=E \ ; \qquad\quad \left(r^{2}+r_{min}^{2}\right)\left(\frac{d\phi}{d\lambda}\right)=L \ .
\end{equation}
Substituting those relations in in Eq. \ref{geodesic}, we find
\begin{equation}
\frac{1}{2} \left(\frac{dr}{d\lambda}\right)^{2} + \frac{1}{2} \left(1 - \frac{2}{\sqrt{r^{2}+r_{min}^{2}}}\right) \left(\frac{L^{2}}{r^{2}+r_{min}^{2}} + 1\right) \ = \frac{E^{2}}{2} \ ,
\end{equation}
which is the analog of Eq.(\ref{eq:GeoSch}) for the Visser-Simpson spacetime. 
So exactly as in the Schwarzschild case, we found that the radial motion is the same as that of a nonrelativistic unit mass particle of energy $E^2/2$, but now moving under the effective potential
\begin{equation}
V_{\rm eff} = \frac{1}{2} -\frac{1}{\left(r^{2}+r_{min}^2\right)^{1/2}}+\frac{L^2}{2\left(r^{2}+r_{min}^2\right)}-\frac{L^2}{\left(r^{2}+r_{min}^2\right)^{3/2}}.
\end{equation}

\begin{figure}[h]
\centering
 \includegraphics[width=0.8\columnwidth]{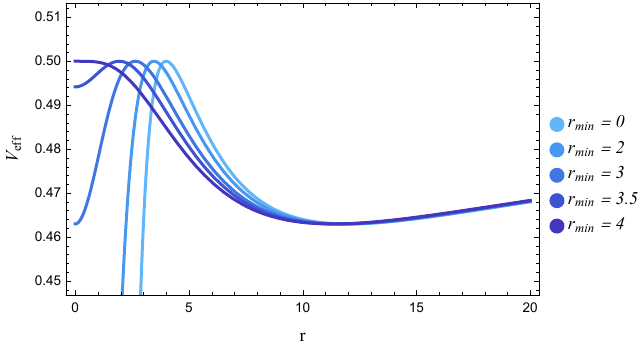}
 \includegraphics[width=0.8\columnwidth]{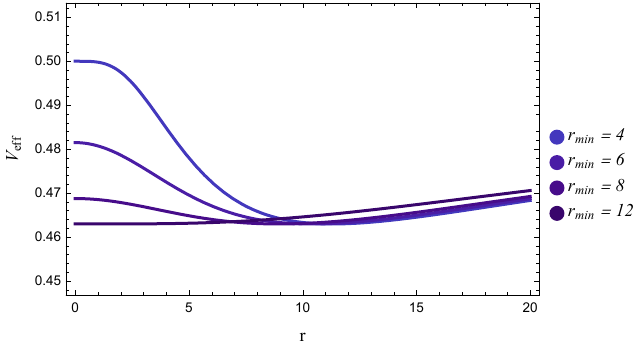}
 \caption{Behavior of the effective potential for the Simpson-Visser metric with the increase of the parameter $r_{min}$ for $L = L_{IBCO} = 4$. Upper plot: $r_{min}$ ranging from $0$ to $4$. Lower plot: $r_{min}$ ranging from $4$ to $12$. Notice that when $r_{min} = 4$ the location of the IBCO is shifted to the origin. For $r_{min} = 12$, it is the radius of the stable circular orbit (for this value of $L$) that is shifted to the origin.}
 \label{fig:Veff_vs}
\end{figure}

If $r_{min} = 0$, of course, this is the same as the effective potential in the Schwarzschild case (eq. \ref{veff_schw}), which we will denote as $V_0(r)$. Then note that the $V_{\rm eff}(r)$ above is the same as $V_0\left(\sqrt{r^{2}+r_{min}^{2}}\right)$. Hence, naively, we could expect that the introduction of $r_{min}$ will cause an effect similar to that of $x \xrightarrow{} x + x_0$, which is a shift to the left (since a given value of $r$ in $V_{\rm eff}$ corresponds to what would be a larger value of $r$ in $V_0$). Note, of course, that this is just a qualitative statement % cannot be the only effect. In fact, $V_{\rm eff}(r)$ is always an even function, thus the symmetry over the y axis has to be maintained, and this translation cannot be directly done. Also, 
because $r_{min}$ causes more significant effects when $r$ is closer to $r=0$ than when $r\gg r_{min}$. %(for instance, if $r_{min} = 3$, then  $\sqrt{4^2 + r_{min}^2} = 5$, dislocating $5 - 4 = 1$ and $\sqrt{12^2 + r_{min}^2} = 3\sqrt{17}$, dislocating $3\sqrt{17} - 12 \approx 0.37$).
To visualize the effect of $r_{min}$ in $V_{eff}$ for a given $L$, see Fig. \ref{fig:Veff_vs}. 
%Then, to see visually what must happen we can imagine that as $r_{min}$ increases the function 'enters' in the y axis (from the left and from the right) and also stretches for lower values of $r$, while maintaining itself even. Figure \ref{fig:Veff_vs} shows this evolution.

To gain some insight on the modifications with respect to the Schwarzschild case, consider an orbit with a given $L$ that starts at a given apastron $r=r_a$ when $r_{min}>0$. The potential that this particle feels corresponds to the potential of a larger apastron $r_a^S$ in the Schwarzschild case (when $r_{min}=0$) because $(r_a^S)^2=r_{min}^2+r_a^2$. Since we are interested in comparing orbits that initiate at the same (physical) radius in both the Schwarzschild and Simpson-Visser models, it is more appropriate to identify the physical apastron as the radius of the 2-spheres, $r_{physical}=\sqrt{r_{min}^2+r^2}$, rather than simply as the value of the radial coordinate $r$. Consequently, in the 'shifted' potential (with $r_{min} \neq 0$), we will adopt the rule that the initial condition must also be shifted to the left, making $r_0 = \sqrt{r_{physical}^2 - r_{min}^2}$), which leads to the same values for energy and angular momentum as the ones we have in the Schwarzschild case for $r=r_0$.\\

In spite of this, it is possible to show that trajectories with increasing $r_{min}$ in the Simpson-Visser model have larger associated rationals than Schwarzschild ones that start at the same physical radius. This can be directly seen by analyzing the accumulated angle between successive apastra $\Delta \varphi _r$. Recalling eq. \ref{eq:delta_phi} and expressing $\Delta\varphi_r^{Sch}=\int_{r_p}^{r_a}\Phi(r)dr$, the corresponding $\Delta\varphi_r^{VS}$ can be written with a simple change of variables $\rho=\sqrt{r_{min}^2+r^2}$ as 
\begin{equation}
\Delta \varphi_r^{VS} = \int_{r_p}^{r_a} \frac{\rho}{\sqrt{\rho^2 - r_{min}^2}}\Phi(\rho) d\rho \ ,
\end{equation}
Since the term $\rho/\sqrt{\rho^2 - r_{min}^2}$ is always greater than one, its effect is basically to increase the value of the integral, causing the azimuth swept by the orbit to increase with $r_{min}$ and, consequently, also increases the associated rational $q$.\\

In this context, it is also immediate to see that the Simpson-Visser model preserves features of the Schwarzschild geometry, as having a lower limit for $q$ represented by the stable circular orbit, which has the lowest energy, and having the upper limit represented by the homoclinic orbit (or $E = 1$ if the maximum in the potential is not bounded).\\

\begin{figure}[!h]
\centering
 \includegraphics[width=1\columnwidth]{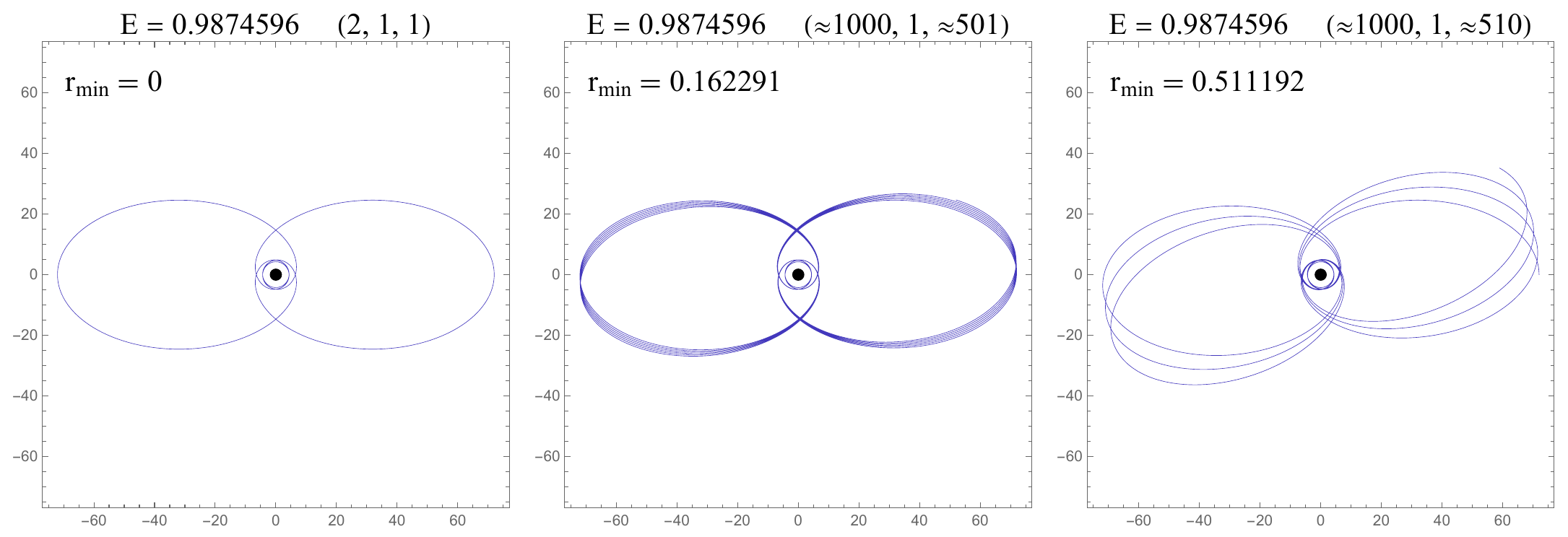}
 \caption{2-leaf orbits in Simpson-Visser metric with increasing $r_{min}$ from left to right. Note that the orbits are just precessions of the original 2-leaf orbit. All these orbits have the same initial physical radius ($r_0 \approx 72.0054$) and angular momemtum $L = 3.9$.}
 \label{fig:2-leaf_precession}
\end{figure}

\begin{figure}[!h]
\centering
 \includegraphics[width=1\columnwidth]{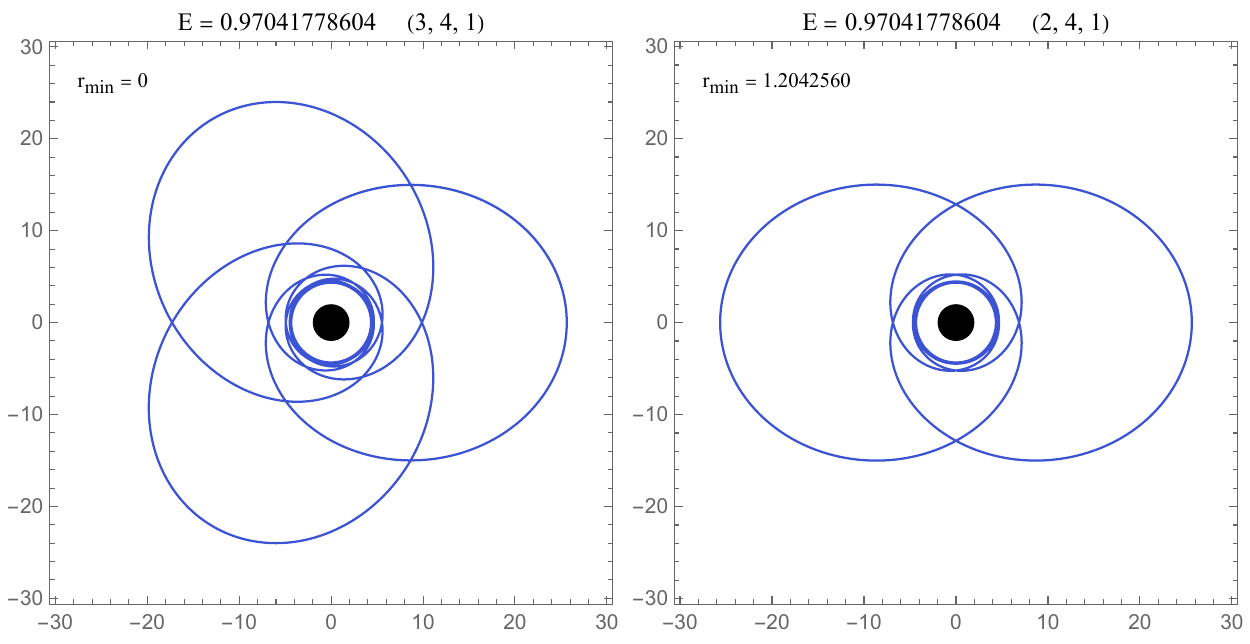}
 \caption{Orbits with same initial radius ($r_0 \approx 25.606327$) but with different values of $r_{min}$ with $L = 3.75$. Left: 3-leaf orbit in Schwarzschild black hole. Right: Increasing $r_{min}$ for the same initial radius the orbit has now two leaves. Notice that because we have more whirls, $q$ is in the steepest part of the curves represented by Figure \ref{fig:qcurves} so the associated rational for orbits with different $r_{min}$ vary more, in contrast with Figure \ref{fig:2-leaf_precession}.}
 \label{fig:3-leaf_to_2}
\end{figure}

On the other hand, recalling that $q = w + v/z$, it is expected that a small increase in $r_{min}$ would not change the number of whirls $w$ of the orbits, because that  would cause an abrupt change in $q$, but only the fractional part $v/z$ (unless, of course, $v/z$ is close to one or zero). This means that we expect to see only small perturbations and, for instance, that we would associate a 2-leaf orbit with $r_{min} = 0$ to a precessing 2-leaf orbit when $r_{min}$ is slightly increased from zero. This is exactly what we show in Fig. \ref{fig:2-leaf_precession}. Also, because $q$ in the Schwarzschild case has the behavior presented in Fig. \ref{fig:qcurves}, if the orbit is in that region where a small increase in the energy causes an abrupt change in $q$, we would not only need to associate it with precessions of the original orbit but also, for instance, relate a Schwarzschild 3-leaf to a Simpson-Visser 2-leaf, as in Fig. \ref{fig:3-leaf_to_2}, causing a significant change in the orbit.\\

With this line of thought, if we increase $r_{min}$ sufficiently slowly (and since $L$, just as in any axisymmetric spacetimes, is an adiabatic invariant \cite{hughes2019bound}) the orbits would continuously evolve describing a family of trajectories with larger and larger values for the associated rational $q$ (because a precessing orbit is simply an orbit with the same number of whirls $w$ but with many more leaves $z$).\\

\begin{figure}[htbp]
\centering
 \includegraphics[height=\dimexpr\textheight-70pt\relax]{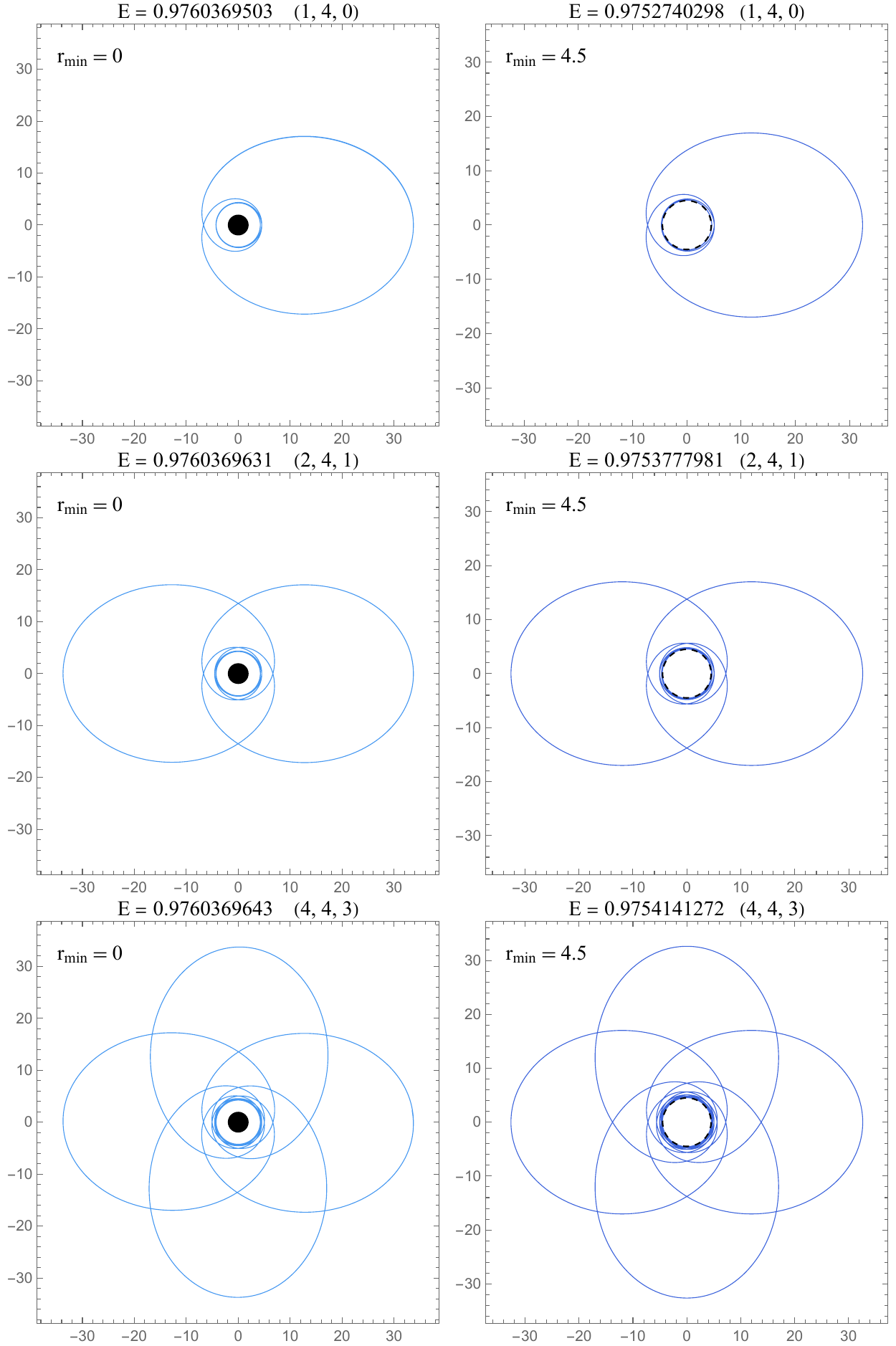}
 \caption{Comparison between orbits in the Schwarzschild black hole (left) and Simpson-Visser black bounce (right) for the same value of angular momentum and associated rational. For the Simpson-Visser case, $r_{min} = 4.5$ (for this value the metric describes a wormhole and the dashed line represents its throat at $r=0$). All orbits have $L = 3.8$ and note that we showed more digits for the values of energy $E$ since in this region with large values of $q$ the associated rational varies abruptly with the energy (see Figure \ref{fig:qcurves}).}
 \label{fig:Schwrd_vs_comp}
\end{figure}

Thus, we have shown that the orbits in the Simpson-Visser model can be reduced to a family with the same topological features as the orbits in the Schwarzschild metric (with the triplet $(z, w, v)$). Additionally, we can also use the translation argument (shift of the potential) to compare the evolution of orbits allowed for given values of angular momentum in Simpson-Visser and Schwarzschild metrics.\\

In this sense, we already know the general behavior for the Schwarzschild case: in the initial region where $L_{ISCO} < L < L_{IBCO}$ orbits can have arbitrarily large $q$ and, as the angular momentum increases, these orbits will have greater and greater eccentricities. Then, when $L > L_{IBCO}$ these high $q$ orbits cease to be allowed, and the upper limit will be the rational associated to the bound orbit that comes from infinity and has the periastron defined by the radius at which $V_{eff} = 1/2$ (or $E=1$, the maximum energy for a bound orbit. See the first two larger $L$ in Figure \ref{fig:Veff}).  In addition, in the weak-field regime (large angular momentum) all orbits cease to exhibit whirls ($w = 0$) and can be described as minute precessions of an ellipse (with a very low ratio $v/z$) as in the case of Mercury's orbit in our solar system.\\

\begin{figure}[htbp]
\centering
 \includegraphics[width=0.8\columnwidth]{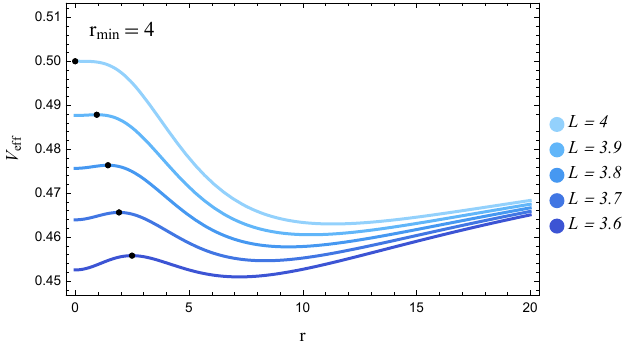}
 \caption{Simpson-Visser effective potential for $r_{min} = 4$, chosen such that $r_{IBCO} < r_{min} < r_{ISCO}$, shown for several values of angular momentum $L$. As $L$ decreases, the local maximum shifts to larger radii, so that orbits associated with arbitrarily large rationals can have physical periastra larger than $r_{min}$.}
 \label{fig:Veff_max}
\end{figure}

Moreover, for small values of $r_{min}$, all these characteristics of the effective potential are preserved, and orbits in Simpson-Visser model will exhibit in general the same behavior as the Schwarzschild case discussed above, with the difference that the periastra become closer to the minimal sphere of area $4\pi r_{min}^2$. This is so until $r_{min} = r_{IBCO}$, because for this value of $r_{min}$ the innermost bound maximum is displaced to the coordinate center $r=0$ (also, at this point, $r_{min} > 2M$ and the spacetime now describes a wormhole). However, for $L < L_{IBCO}$ this maximum is offset further to the right and would not be totally displaced, as illustrated in Fig. \ref{fig:Veff_max}. Still, for larger values of $r_{min}$, these maxima are progressively reached, requiring smaller and smaller values of $L$ to avoid having a maximum displaced to $r=0$.

However, this maximum cannot be indefinitely shifted to the right. In fact, the farthest it can go is when $r = r_{ISCO}$ (the point at which the unstable circular orbit merges with the stable one). This establishes another region, where $r_{IBCO} < r_{min} < r_{ISCO}$. Note also that as we increase $r_{min}$ and these maxima are reached, because this range has values for $V_{\rm eff}$ greater than at the minimum (obviously), and recalling that $V_{\rm eff}$ is even, they necessarily become maxima at the origin (see the lower plot in Figure \ref{fig:Veff_vs}). For this reason,  it is possible to associate homoclinic orbits (which would have a physical periastron approaching $r_{min}$) with them, causing orbits with arbitrarily large $q$ to exist within this range of angular momenta.\\ 

Therefore, in this case, just as in the Schwarzschild description, there is also a range of angular momentum within which orbits have arbitrarily large associated rationals, until larger values of $L$, where $q$ becomes bounded. The essential difference is that beyond a certain apastron radius, the trajectories with arbitrarily large associated rationals will also have physical periastra arbitrarily close to $r_{min}$ (since the maximum will be in the origin for theses cases). Figure \ref{fig:Schwrd_vs_comp} illustrates this by comparing Schwarzschild and Simpson-Visser orbits in this intermediate case with the same value for $L$ and $q$.\\

It is important to point out that, despite the orbits in the Schwarzschild and Simpson-Visser space-times looking similar (since they have the same associated triplet), they are the result of integrating completely different equations of motion. In order to observationally tell them apart, one would need a substantial degree of precision. If one ignores the nature of the central object, focusing only on the general aspects of the orbital motion (such as the associated triplet), then the orbits would be hardly distinguishable.\\ 

\begin{figure}[htbp]
\centering
 \includegraphics[width=0.8\columnwidth]{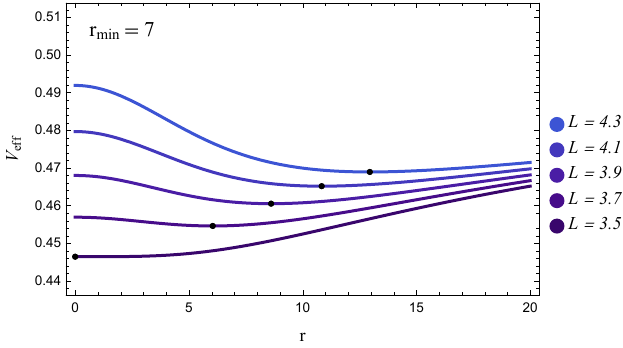}
 \caption{Simpson-Visser effective potential for $r_{min} = 7$, chosen such that $r_{min} > r_{ISCO}$, shown for several values of angular momentum $L$. As $L$ increases, the location of the potential minimum shifts to larger radii. In this case, however, in constrast with Figure \ref{fig:Veff_max}, all orbits associated with arbitrarily large rationals have physical periastra arbitrarily close to $r_{min}$.}
 \label{fig:Veff_min}
\end{figure}

Further, when $r_{min} > r_{ISCO}$ not only will we be prohibiting any maximum from existing in $r > 0$, but also will start shifting some minima to the origin. For these cases, the particle cannot describe trajectories only in the region $r > 0$ because it has to go through the wormhole. As $L$ increases, the radius of the stable circular orbit becomes larger and, for the same reasons as before, a maximum may appear at the throat ($r=0$), as shown in Fig. \ref{fig:Veff_min}. With increasing $L$, this maximum of $V_{\rm eff}$ will grow in amplitude until it becomes larger than $1/2$ (which corresponds to $E = 1$). Thus, in previous cases, for an initial range of angular momenta, orbits will have $q$ arbitrarily large, but they will also have physical periastra arbitrarily close to $r_{min}$ (since the maximum is always at the origin). Additionally, in contrast with the other situations, closed orbits (with $r > 0$ only) will not exist for the same range of angular momentum as in the Schwarzschild case (that is, $L > L_{ISCO}$). As $r_{min}$ increases, one needs a larger minimum angular momentum for this to happen (and it is given by the value of $L$ that turns $r = r_{min}$ into a minimum for $V_0$).\\ %Hasta aqui.

Perhaps another important aspect to note is that this range of $L$ in which high $q$ orbits are allowed is roughly determined by the difference of the angular momentum in which $V_0(r_{min})$ is a minimum and the one for which $V_0(r_{min}) = 0.5$, and that this difference increases monotonically with $r_{min}$. Hence, as $r_{min}$ increases (within this region where $r_{min} > r_{ISCO}$), the range of angular momenta that allows for orbits with arbitrarily large associated rationals also increases. Consequently, we expect such orbits to become more frequent as $r_{min}$ increases.

So far, we have only considered orbits that do not go through the wormhole. Next, we are going to specifically analyze bound orbits that can traverse it. \\ %Hasta aqui.

\subsection{Wormhole case: traversing orbits}
\label{subsec:trav_worm}

\begin{figure}[!h]
\centering
 \includegraphics[width=1\columnwidth]{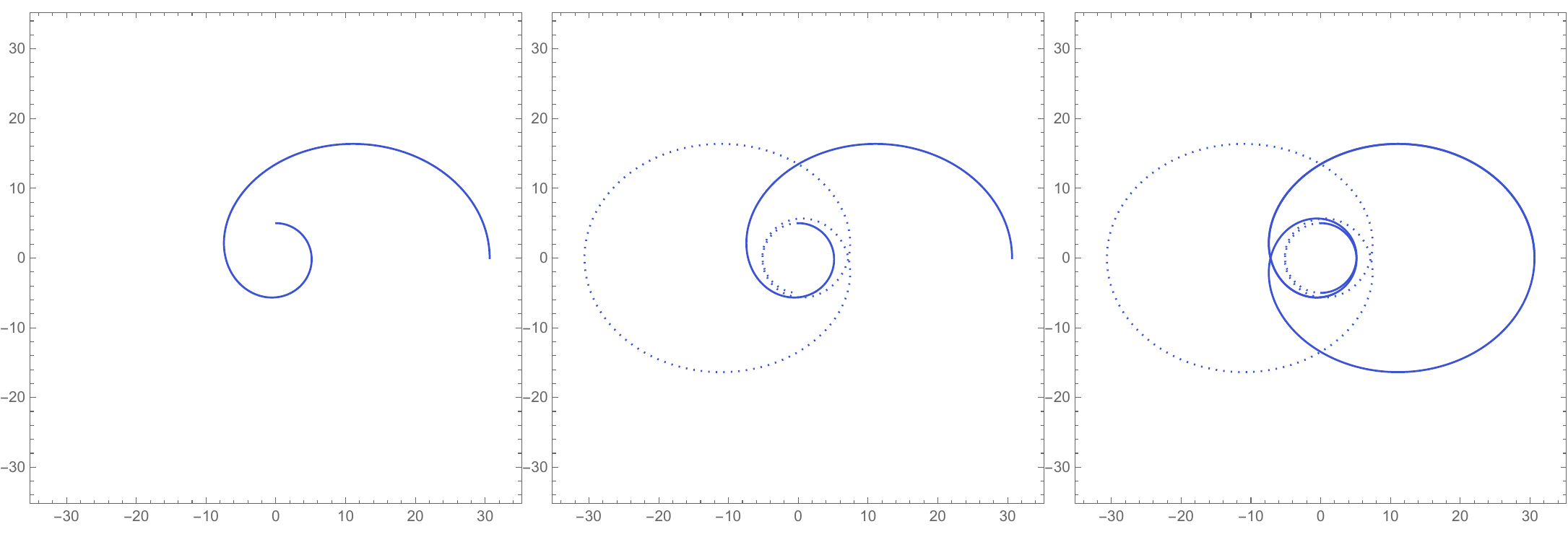}
 \caption{Example of an orbit with observed triplet $(1,1,0)$ that comes from $(2,1,1)$. Each plot shows the orbit before the next throat-crossing, and the dotted part represents the trajectory in $r < 0$. This orbit has $r_{min} = 5$, $E = 0.9740256$ and $L = 3.75$.}
 \label{fig:vs_wormhole_2-leaf}
\end{figure}

As noted before, if $r_{min} > 2M$ the Simpson-Visser metric describes a traversable wormhole geometry. In this case the wormhole throat is described by the hypersurface at $r = 0$, with $r < 0$ corresponding to the universe on the other side of the geometry \cite{simpson2019black}.

Like in the previous case, the analysis of the equations of motion and the effective potential will be very helpful. As noted before, since the equations are symmetric under the exchange $r \xrightarrow{} -r$, the trajectory followed by a test particle on the $r > 0$ part of the spacetime is completely analogous to the trajectory in the $r < 0$ region. This allows us to represent the trajectory in the $r < 0$ region as a continuation of the orbit in $r > 0$. This implies that it is possible to use the integers triplet $(z, w, v)$ to describe bound orbits in this case.

\begin{figure}[!h]
\centering
 \includegraphics[width=1\columnwidth]{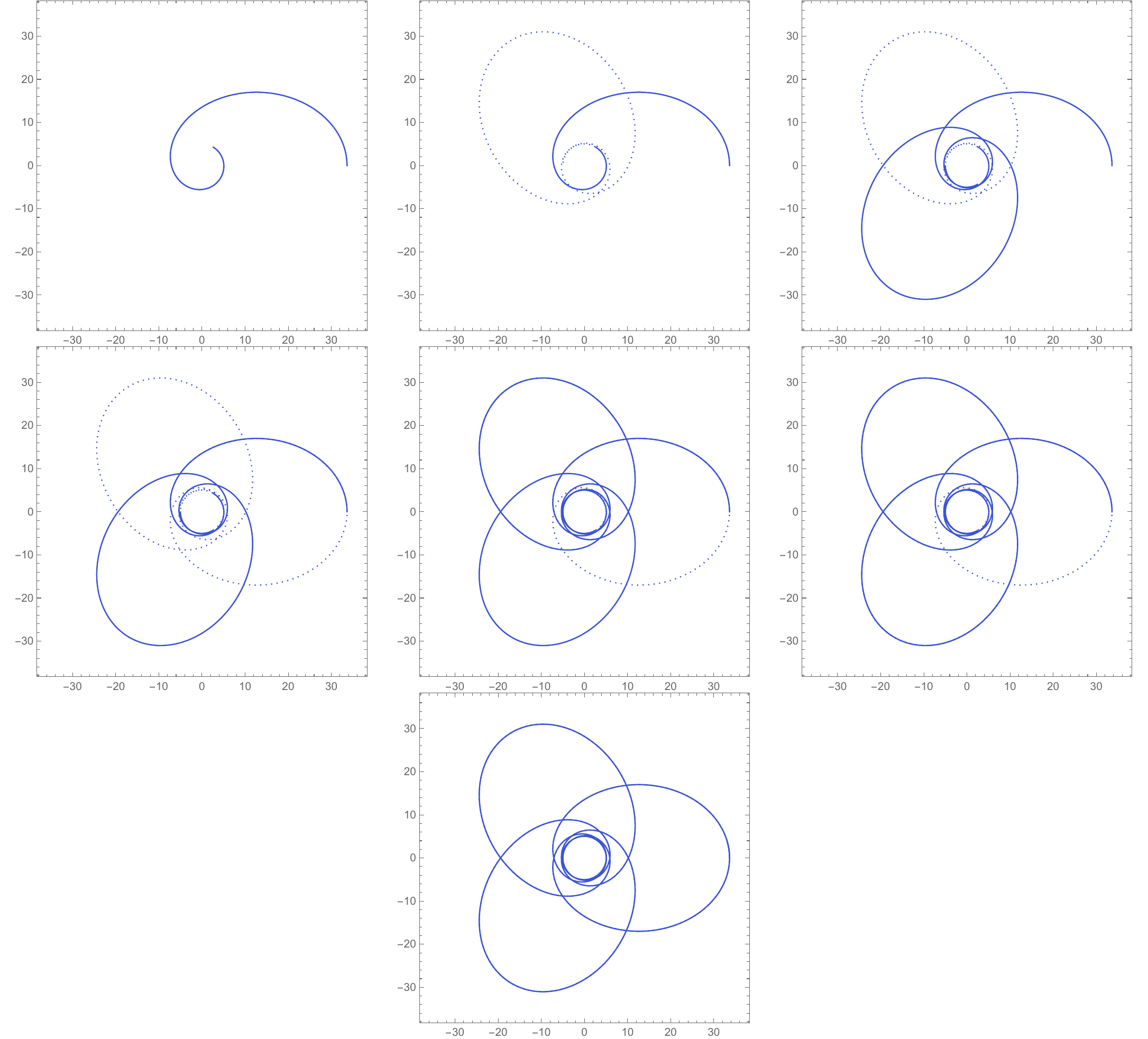}
 \caption{Example of an orbit with observed triplet $(3,1,2)$ that comes from $(3,1,1)$. Each plot shows the orbit before the next throat-crossing, and the dotted part represents the trajectory in $r < 0$. This orbit has $r_{min} = 5$, $E = 0.9758229$ and $L = 3.75$.}
 \label{fig:vs_wormhole_3-leaf}
\end{figure}

In this context, we proceed to analyze how those orbits would be represented. Firstly, if the orbit has a given number $w$, the trajectory will make $w$ whirls in the $r < 0$ region before returning to the $r > 0$ region. In addition, it will also complete one of the $z$ leaves (skipping $v$ vertices, as already discussed) and accumulate an angle equal to $2\pi v/z$. In this sense, though this part of the trajectory may not be observable from the $r>0$ region\footnote{A dedicated study of the propagation of light rays emitted by a luminous point source \cite{Rosa:2023qcv} in this geometry would be necessary to assess the observability of these orbits, but that lies beyond the scope of this paper and will be considered elsewhere. }, the accumulated angle still contributes. For this reason, for an observer in our universe, the orbit will sweep an angle equivalent to $w$ whirls plus an angle of $2\pi v/z$. If $2v/z$ is larger than one, this contribution would cause one more apparent whirl, and the remainder is the new observed vertex-leaf relation. Otherwise, the observed trajectory will have a new number $v_{obs}$ equal to $2v$, and if it is exactly equal to one (that is, a two-leaf) it will be visible as a one-leaf orbit.\\

So, in short, given the rational number $q$ related to the accumulated angle between successive apastra by eq. \ref{dphir} (considering the $r < 0$ trajectory),  the observable orbit, which is detected by an observer with $r > 0$, has a triplet $(z, w, v)_{obs}$ corresponding to the rational
\begin{equation}
    q_{obs} = \lfloor q \rfloor + 2(q - \lfloor q \rfloor) = 2q - \lfloor q \rfloor
\end{equation}

We have thus established a classification scheme for orbits that traverse the wormhole that is analogous to that used in non-wormhole space-times. As illustrative examples, Fig. \ref{fig:vs_wormhole_2-leaf} shows an orbit originating from the triplet $(2,1,1)$ but observed as having one whirl and one leaf, while Fig. \ref{fig:vs_wormhole_3-leaf} presents a trajectory associated with $(3,1,1)$ but observed as $(3,1,2)$ (all in accordance with the explicitly stated rule).

\begin{figure}[!h]
\centering
 \includegraphics[width=0.8\columnwidth]{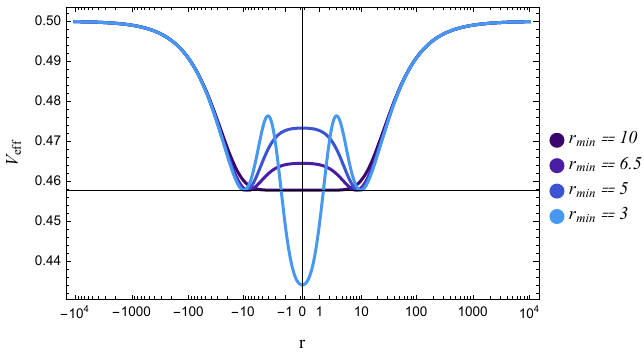}
 \caption{Instances of the effective potential for various values of $r_{min}$ in cases where bound traversing orbits are possible. These cases have $L = 3.8$ and are plotted in a log $r$ scale (and are replicated for $r < 0$ since the function for $V_{eff}$ is even).}
 \label{fig:vs_Veff_wormhole}
\end{figure}

Furthermore, from the effective potential, one concludes that it is possible to have two distinct regions where the traversing orbits are essentially different. Note from Fig. \ref{fig:vs_Veff_wormhole}, where the effective potential is shown for some values of $r_{min}$, that as $r_{min}$ is increased, at first ($r_{min} = 3$ in the figure) bound traversing orbits exist for radii ranging from zero to the radius corresponding to the local maximum of the effective potential (that we will call inner orbits) and for trajectories with energy greater than the energy of this local maximum (referred to as outer orbits). Additionally, for $r_{min}$ sufficiently large (greater than the unstable circular orbit radius when $r_{min} = 0$) inner orbits cease to exist, without any distinction of regions for orbits that enter the wormhole.

\begin{figure}[!h]
\centering
 \includegraphics[width=1\columnwidth]{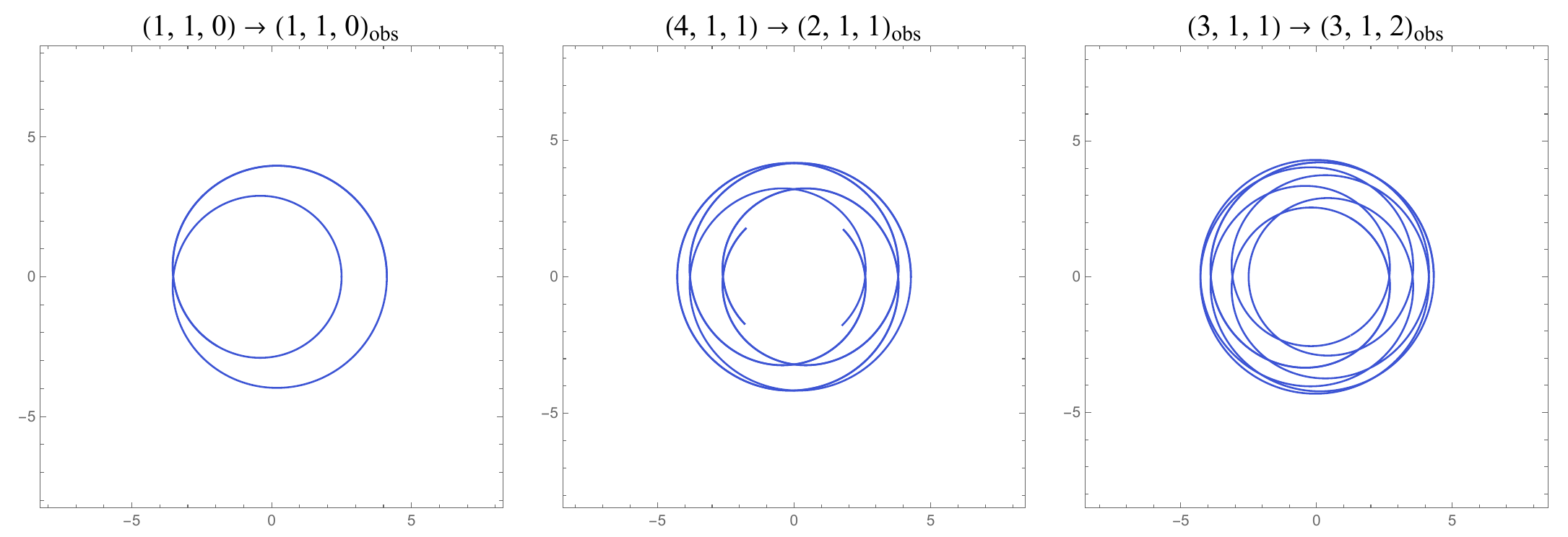}
 \caption{Examples of traversing orbits in the inner part of the effective potential (note the small eccentricities). In the top of each plot the triplet associated with the accumulated angle $\Delta \varphi _r$ is indicated, together with the triplet that is actually observed in $r>0$. All trajectories have $r_{min} = 2.5$ and $L = 3.6$. The energies are, from left to right, $E = 0.9525868$, $E = 0.9537071$, and $E = 0.9539270$.}
 \label{fig:vs_wormhole_inner_orbits}
\end{figure}

We also note that the values of the associated rationals have a different relation with the energy (or the eccentricity) for inner and outer orbits. For outer orbits, the lower limit for the energy is determined by the local maximum and the upper limit by the value at infinity ($E = 1$). However, we know that, as the orbit approaches the maximum, the associated rational approaches infinity (because we can associate to it an orbit similar to a homoclinic one) and, as the initial conditions move away from the local maximum, the orbits will whirl less and less and approach the upper limit ($E = 1$) where it describes an orbit with some finite $q$. With this in mind, we see that, contrary to the behavior in previous cases, the associated rational decreases with increasing energy. This line of thought, of course, is still valid for those cases in which the local maximum is at the origin.

Nonetheless, for inner orbits the reasoning and limits are similar to the no-through trajectories already discussed, and the associated rational increases with the energy. So, in summary, for outer orbits, unlike other cases, the associated rational $q$ monotonically decreases with energy (or eccentricity), whereas it increases for inner orbits. Another distinctive aspect is that inner orbits have smaller eccentricities compared with outer ones. Further, for the cases where the local maximum is at the origin the description is similar to that of the outer orbits examples. For the instances where the origin corresponds to a minimum (and there are no local maxima), both limits become finite, defined by the value that $q$ takes as the orbit approaches infinity and as it approaches the stable circular orbit at the origin. Figs. \ref{fig:vs_wormhole_2-leaf} and \ref{fig:vs_wormhole_3-leaf} are examples of outer orbits, and Fig. \ref{fig:vs_wormhole_inner_orbits} presents some inner orbits.

\section{Equatorial orbits in the rotating Simpson-Visser metric}

Just like the Kerr metric, which, despite not admitting a simple description through a one-dimensional effective potential, has orbits with analogous descriptions as in the Schwarzschild case, the equatorial geodesics in the Simpson-Visser metric with rotation are completely analogous to those we discussed in the previous section. To see this, let us first explain how to derive the corresponding rotating solution.\\

The general parametrization of a static, spherically symmetric metric is
\begin{equation}
 ds^2 = - f(r)dt^2 + \frac{dr^2}{f(r)} + h(r) [d\theta^2 + \sin^2 \theta d\phi^2];
\label{sphr_symm_met}
\end{equation}
A common way employed for finding the axisymmetric counterpart of \ref{sphr_symm_met} is the Newman-Janis algorithm \cite{newman1965note, newman1965metric, afonso2022infinite}, consisting, basically, in a particular complexification of radial and time coordinates with a complex coordinate transformation (see \cite{erbin2017janis} for a review). Several other prescriptions of this method can be used. As an example, in \cite{azreg2014generating, azreg2014static} the complexification is substituted by some physical arguments and symmetry properties, naturally generating solutions in Boyer-Lindquist coordinates. In our case, we have
\begin{equation}
f(r) = 1-\frac{2M}{\sqrt{r^2 + r_{min}^2}}, \quad h(r) = r^2 + r_{min}^2.
\end{equation}
With this choice, the metric obtained by the Newman-Janis procedure can be written, in Boyer-Lindquist coordinates, as (see \cite{mazza2021novel}):
\begin{equation}
    ds^2 = -\left(1 - \frac{2M \sqrt{r^2 + r_{min}^2}}{\Sigma}\right) dt^2 + \frac{\Sigma}{\Delta} \, dr^2 + \Sigma \, d\theta^2 - \frac{4M a \sin^2 \theta \sqrt{r^2 + r_{min}^2}}{\Sigma} \, dt d\varphi 
    + A \sin^2 \theta \, d\varphi^2
    \label{rotSV_met}
\end{equation}
where $a$ is the usual spin parameter and $M$ the total mass, with
\begin{gather*}
\Sigma^2 = r^2 + r_{min}^2 + a^2 \cos^2 \theta, \qquad \Delta = r^2 + r_{min}^2 - 2M\sqrt{r^2 + r_{min}^2} + a^2, \\
A = r^2 + r_{min}^2 + a^2 + \frac{2M a^2 \sin^2 \theta \sqrt{r^2 + r_{min}^2} }{\Sigma}
\end{gather*}
As one can see, the above line element reduces to the Kerr metric when $r_{min} = 0$ but $a\neq 0$, and when $a = 0$ but $r_{min}\neq 0$ it becomes the Simpson-Visser metric. Actually, analogously to the relation between the SV model and the Schwarzschild metric, this metric is the Kerr metric exchanging $r$ by $\sqrt{r^2 + r_{min}^2}$, but without changing $dr$ (thus, this metric is not related to Kerr by a change of coordinates). Also, depending on the values of $a$ and $r_{min}$ this metric can represent different kinds of regular black holes or wormholes (see \cite{mazza2021novel} for a more detailed discussion).\\

To derive the equations for the geodesics, we can use the Hamilton-Jacobi method \cite{chandrasekhar1998mathematical}. The Hamilton-Jacobi equation governing geodesic motion with a metric $g^{\mu\nu}$ is 
\begin{equation}
\frac{\partial S}{\partial \tau} = -\frac{1}{2}g^{\mu\nu}\frac{\partial S}{\partial x^\mu}\frac{\partial S}{\partial x^\nu} \ .
\label{ham-jac}
\end{equation}
Assuming that the variables are separable, the solution must be of the form
\begin{equation}
    S = \frac{1}{2}\,\mu^2 \tau - Et + L\varphi + S_r(r) + S_\theta(\theta) \ ,
\end{equation}
where $E$ and $L$ represent, as usual, the energy per unit mass and the angular momentum per unit mass, respectively, while $\mu^2 = 0$ for null geodesics and $\mu^2 = 1$ for time-like geodesics (which is our case). Substituting this form of $S$ in equation \ref{ham-jac} we find a set of equations which is analogous to the Kerr case, namely, 
\begin{subequations}
\begin{align}
        \Sigma \frac{dt}{d\tau} &= a(L - aE\sin^2\theta) + \frac{r^2 + r_{min}^2 + a^2}{\Delta}[E(r^2 + r_{min}^2 + a^2) - La],\\
        \Sigma \frac{dr}{d\tau} &= \pm \sqrt{R}, \label{r_eq}\\
        \Sigma \frac{d\theta}{d\tau} &= \pm \sqrt{\Theta}, \label{theta_eq}\\
        \Sigma \frac{d\varphi}{d\tau} &= \frac{L}{\sin^2\theta} - aE + \frac{a}{\Delta} [E(r^2 + r_{min}^2 + a^2) - La],
\end{align}
\label{vs_with_rot_eqs}
\end{subequations}
with
\begin{gather*}
R = [E (r^2 + r_{min}^2 + a^2) - La]^2 - \Delta[\mu^2(r^2 + r_{min}^2) + (L -aE)^2 + Q],\\
\Theta = Q - \cos^2\theta \left[a^2 (\mu^2 - E^2) + \frac{L^2}{\sin^2 \theta} \right]
\end{gather*}
where $Q$ has the same expression as the Carter constant of Kerr geodesics:
\begin{equation}
    Q = u_\theta^2 + \cos^2\theta \left[a^2 (1-E)^2 - \frac{L}{\sin^2\theta} \right].
\end{equation}
From now on, we will consider, as in previous sections, $M = 1= \mu^2$, with all the Boyer-Lindquist coordinates interpreted as dimensionless quantities. It is important to point out that these equations of motion are not the optimal choice for computational simulations, since, for example, the need to change signs at the turning points in the equation for $r$ naturally accumulates significant errors. For this reason, as we will see, the use of the Hamiltonian formalism is more suitable \cite{levin2008}.
Let us first consider the Lagrangian for a free particle:
\begin{equation}
{\cal L} = \frac{1}{2} g_{\alpha \beta} \dot{q^\alpha} \dot{q^\beta}.
\label{lagrangian}
\end{equation}
In our case (since $\mu = 1$ and $g_{\alpha \beta} \dot{q^\alpha} \dot{q^\beta} = -1$ along a timelike trajectory) this implies that $\cal L$ is identically equal to $-1/2$. Moreover, the particle's 4-momentum is also dimensionless and identical do the 4-velocity, $p^\alpha \equiv \dot{q}^\alpha$. Defining the canonical momentum $p_\alpha$, we have:
\begin{equation}
    p_\alpha \equiv \frac{\partial {\cal L}}{\partial\dot{q}^\alpha} = g_{\alpha \beta} \dot{q}^\beta = g_{\alpha \beta} p^\beta.
\end{equation}
Hence, explicitly, the components are:
\begin{subequations}
\begin{align}
  p_t &= -\left(1-\frac{2\sqrt{r^2 + r_{min}^2}}{\Sigma}\right ) \dot t -\frac{2a \sqrt{r^2 + r_{min}^2} \sin^2\theta}{\Sigma}\dot \varphi, \label{e} \\
  p_r &= \frac{\Sigma}{\Delta} \dot r, \label{pr}\\
  p_\theta &= \Sigma \dot{\theta}, \label{ptheta}\\
  p_\varphi &= \sin^2\theta\left
  (r^2 + r_{min}^2 + a^2 + \frac{2a^2 \sqrt{r^2 + r_{min}^2} \sin^2\theta}{\Sigma}\right ) \dot\varphi -\frac{2a \sqrt{r^2 + r_{min}^2} \sin^2\theta}{\Sigma}\dot t.\label{pphi}
\end{align}
\end{subequations}
With this we define the Hamiltonian:
\begin{equation}
H = p_\mu \dot{q}^\mu - {\cal L}
= \frac{1}{2}g^{\alpha \beta} p_\alpha p_\beta.
\end{equation}
Thus, in principle, we just need to compute the inverse metric $g^{\alpha \beta}$ to determine the Hamiltonian and derive the equations of motion. However, in this case, it is simpler to use the equations of motion we already have. Since there are no cross terms involving $r$ or $\theta$ in the line element \ref{rotSV_met}, we can write the Hamiltonian as
\begin{equation}
H(\mathbf{q}, \mathbf{p}) =
\frac{\Delta}{2\Sigma}p_r^2 + \frac{1}{2\Sigma}p_\theta^2 + f(r, \theta, p_t, p_\varphi),
\end{equation}
being $f$ a function to be determined. Now, we can use equations \ref{pr}-\ref{ptheta} together with \ref{r_eq}-\ref{theta_eq} and note that, as the Lagrangian, the Hamiltonian must be equal to $-1/2$:
\begin{equation}
H(\mathbf{q}, \mathbf{p}) =
\frac{R}{2\Delta\Sigma} + \frac{\Theta}{2\Sigma} + f(r, \theta, p_t, p_\varphi) = -\frac{1}{2}
\end{equation}
With that we fix $f$ and write $H$ as:
\begin{equation}
H(\mathbf{q}, \mathbf{p}) =
\frac{\Delta}{2\Sigma}p_r^2+\frac{1}{2\Sigma}p_\theta^2
-\frac{R + \Delta\Theta}{2\Delta \Sigma} -\frac{1}{2},
\end{equation}
interpreting $E$ and $L$ as $-p_t$ and $p_\varphi$, respectively.
From this expression we write Hamilton's equations
\begin{equation}
{\dot q_i} =\frac{\partial H}{\partial p_i} \quad , \quad
{\dot p_i}=-\frac{\partial H}{\partial q_i}
\end{equation}
explicitly as
\begin{subequations}
\begin{align}
    \dot{r} & = \frac{\Delta}{\Sigma}p_{r} \label{ham_eqs_0}  \\
    \dot{p}_{r} & = 
    -\frac{\partial}{\partial r}\left(\frac{\Delta}{2\Sigma}\right )p_{r}^{2} -\frac{\partial}{\partial r}\left(\frac{1}{2\Sigma}\right )p_{\theta}^{2} + \frac{\partial}{\partial r}\left(\frac{R +
    \Delta\Theta}{2\Delta\Sigma}\right )  \\
    \dot{\theta}& =  \frac{1}{\Sigma}p_{\theta} 
     \\
    \dot{p}_{\theta} & = 
    -\frac{\partial}{\partial \theta}\left(\frac{\Delta}{2\Sigma}\right )p_{r}^{2} - \frac{\partial}{\partial \theta}\left(\frac{1}{2\Sigma}\right )p_{\theta}^{2} + \frac{\partial}{\partial \theta}\left (\frac{R +
    \Delta\Theta}{2\Delta\Sigma}\right )   \\
    \dot{t} & =  \frac{1}{2\Delta\Sigma} \frac{\partial}{\partial E} \left(R +
    \Delta\Theta \right) \\ 
    \dot{p_t} & =  0  \\
    \dot{\varphi} & =  -\frac{1}{2\Delta\Sigma}
    \frac{\partial}{\partial L} \left(R + \Delta\Theta \right) \\
    \dot{p_\varphi} & =  0 \label{ham_eqs_f}\, ,
\end{align}
\end{subequations}

These equations are completely analogous to those governing geodesic motion in the Kerr metric (see Appendix A of \cite{levin2008}). Though we now have more equations than before, this representation in terms of eqs. \ref{ham_eqs_0}-\ref{ham_eqs_f} eliminates the issues with numerical precision at turning points of our original set. Moreover, since the equations for $\dot{p_t}$ and $\dot{p_\varphi}$ vanish, we do not need to solve them. Additionally, we do not require the variable $t$ for the analysis that follows and focus exclusively on the equatorial orbits. Consequently, in practice, we only need to solve the three equations for $r$, $p_r$ and $\varphi$.

Moreover, if we try to write the radial motion as in the previous section (in the form of a classical motion with a given effective potential), we would find (in the same way as in the Kerr metric \cite{hobson2006general, jefremov2015innermost}) an `effective potential'  that, awkwardly, also depends on the energy:
\begin{equation}
    V = \frac{1}{2} - \frac{1}{(r^2 + r_{min}^2)^{1/2}} - \frac{a^2(E^2 - 1) - L^2}{2(r^2 + r_{min}^2)} - \frac{(L - aE)^2}{(r^2 + r_{min}^2)^{3/2}}.
\end{equation}

\begin{figure}[h]
\centering
 \includegraphics[width=1\columnwidth]{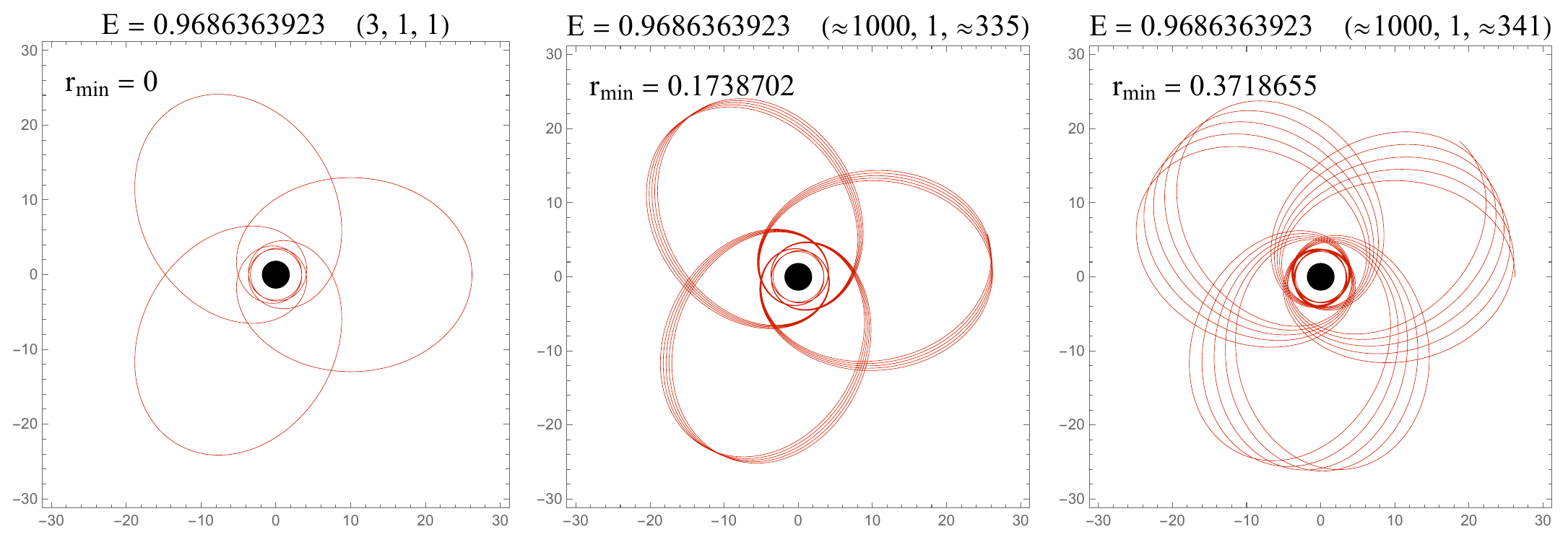}
 \caption{Orbits in the rotating Simpson-Visser metric with increasing $r_{min}$ from left to right causing perturbations in the original $r_{min} = 0$ orbit. All orbits have $L = 3.25$ and $a = 0.5$ and initiate at the same physical radius ($r_0 \approx 26.2180511$) and for this range of $r_{min}$ we have regular black holes, with the black circles representing the event horizons.}
 \label{fig:vs_with_rotation_3-leaf_precession}
\end{figure}

\begin{figure}[h]
\centering
 \includegraphics[width=1\columnwidth]{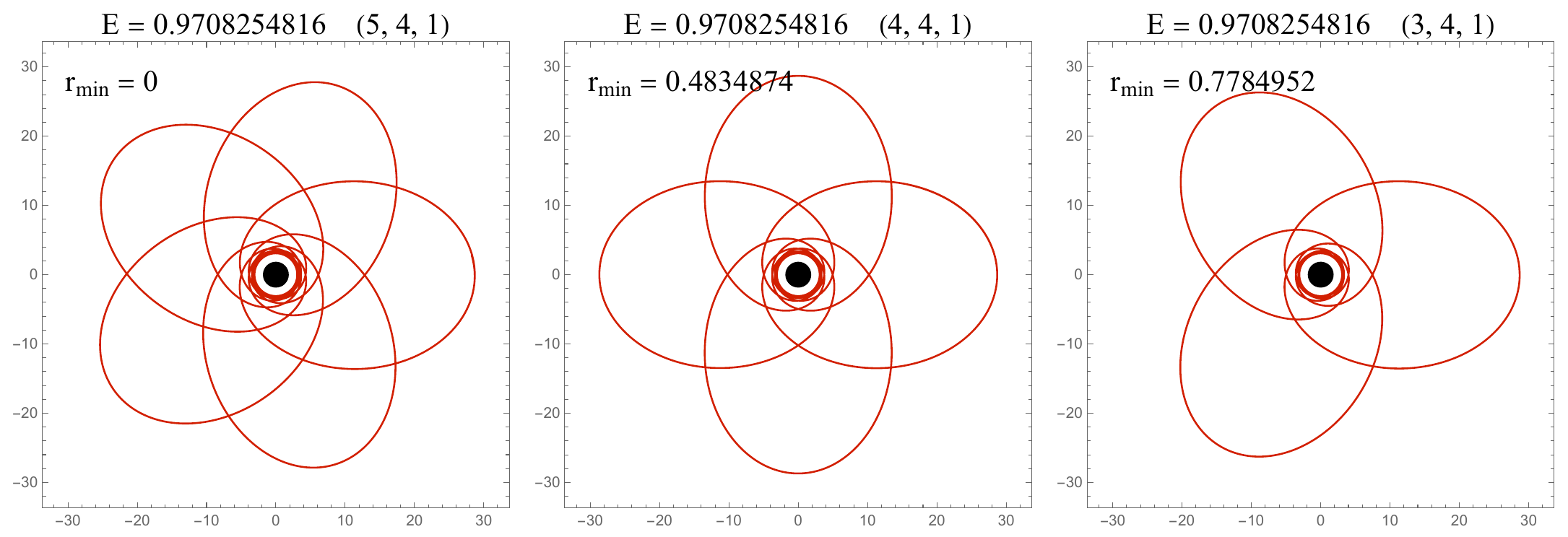}
 \caption{Orbits in the rotating Simpson-Visser metric with increasing $r_{min}$ from left to right causing orbits to change the number of leaves. All orbits have $L = 3.25$ and $a = 0.5$ and initiate at the same physical radius ($r_0 \approx 28.6964379$) and for this range of $r_{min}$ we have regular black holes, with the black circles representing the event horizons.}
 \label{fig:5-leaf_to_4_and_3}
\end{figure}

Nevertheless, the same analogy made between the Schwarzschild and the Simpson-Visser metrics can now be made in the case with rotation. In fact, all the equations of motion in this rotating Visser-Simpson metric are identical to the Kerr case substituting $r$ by $\sqrt{r^2 + r_{min}^2}$ in the way we commented before. Hence, for example, the relation established in the previous section that orbits initiated with the same radius and angular momentum increase the rational number associated with $r_{min}$ is valid here. Thus, in the same way, we expect that with little increases in $r_{min}$ we would see precessions of the orbits with $r_{min} = 0$, but for the orbits with high $q$ (see Fig. \ref{fig:qcurves}), where the associated rational starts to vary abruptly, we would see more significant changes, as commented before. To illustrate this, Fig. \ref{fig:vs_with_rotation_3-leaf_precession} shows the effect of an increase in $r_{min}$ causing little perturbations on a 3-leaf orbit, and in Fig. \ref{fig:5-leaf_to_4_and_3} we show more significant changes in orbits with high associated rationals.

\begin{figure}[!h]
\centering
 \includegraphics[width=1\columnwidth]{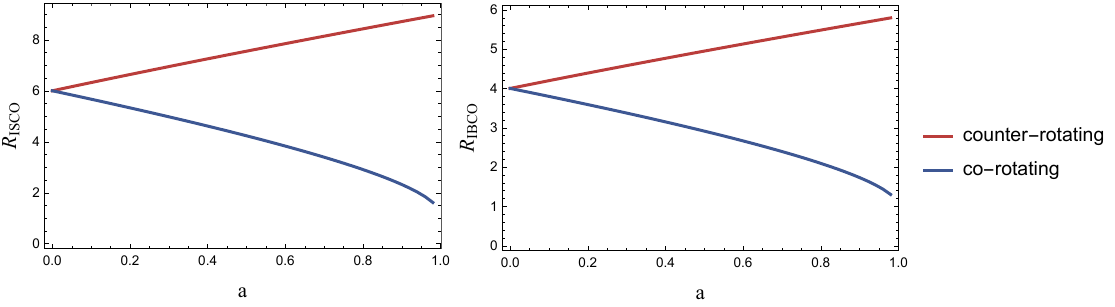}
 \caption{Left: Radius of the innermost stable circular orbit (ISCO) as a function of the spin parameter $a$ for co- and counter-rotating orbits. Right: Radius of the innermost bound circular orbit (IBCO) as a function of the spin parameter $a$ for the co- and counter-rotating orbits.}
 \label{fig:r_isco_ibco}
\end{figure}

\begin{figure}[p]
\centering
 \includegraphics[height=\dimexpr\textheight-70pt\relax]{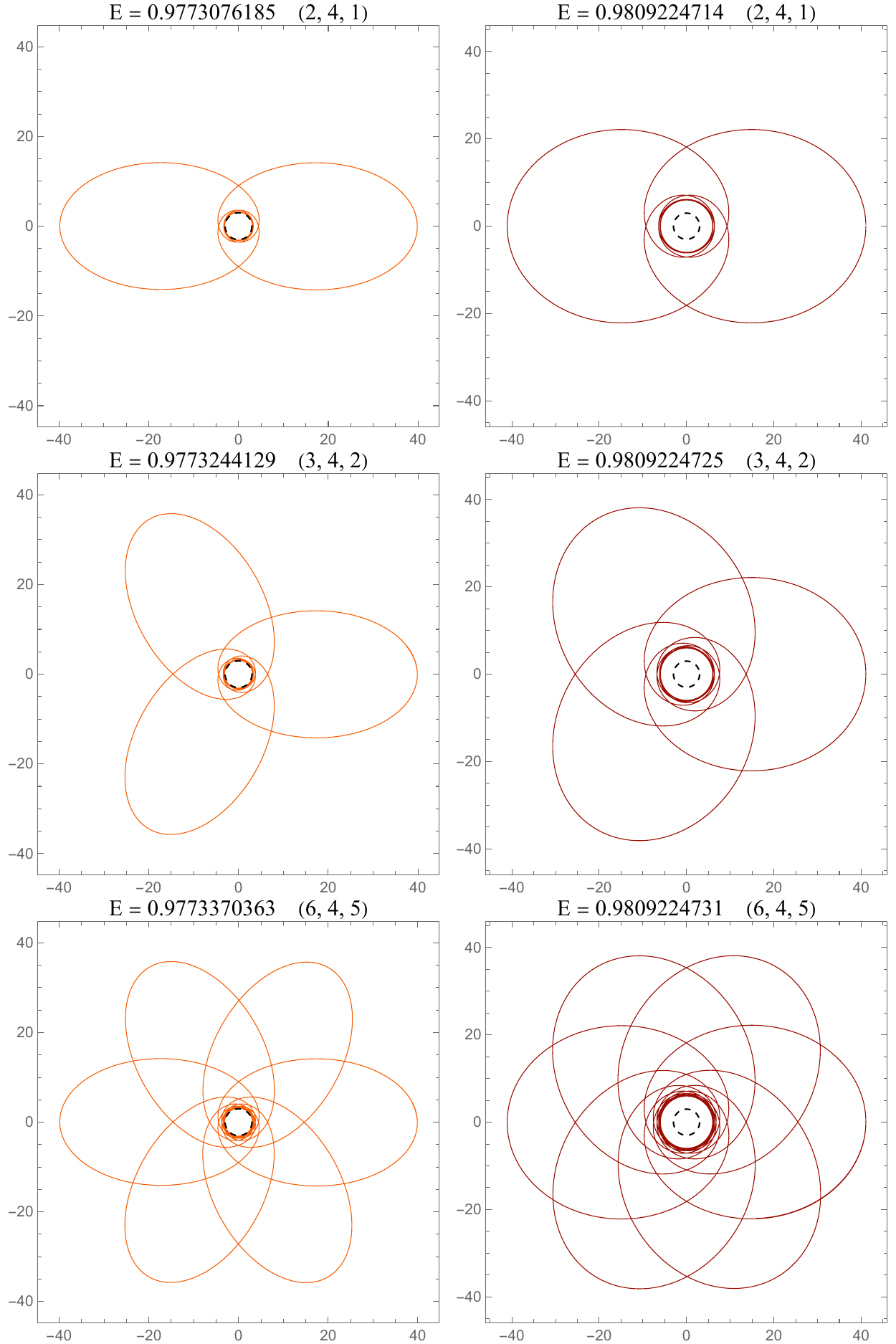}
 \caption{Orbits in the rotating Simpson-Visser metric in a situation where $r_{min}$ is greater than $r_{ISCO}$ for the co-rotating case but less than the $r_{IBCO}$ for the counter-rotating case. The orbits have $a = 0.8$ and $r_{min} = 3$ (for this value of $r_{min}$ the spacetime describes a wormhole). Also, the prograde orbits have $L = 2.95$ and the retrograde $L = -4.44$. The left column shows the co-rotating orbits, and the right column shows counter-rotating orbits with the same triplet $(z,w,v)$.}
 \label{fig:vs_with_rot_co_counter}
\end{figure}

Equatorial orbits in the Kerr metric have the same essential features as the Schwarzschild metric. Specifically, 1) orbits with higher eccentricity have higher values for the associated rational $q$, 2) there is a strong-field regime where $L_{IBCO} < L < L_{ISCO}$ with orbits ranging from the stable circular orbits to the homoclinic one, 3) the associated rational exhibits an upper bound for orbits in the weak-field limit, and 4) all the aspects commented in Sec.\ref{sec:classification}. Consequently, since $r_{min}$ affects orbits in the same way as discussed in the previous section, the orbits in the rotating Simpson-Visser metric have a description analogous to the case without rotation. The main difference is that now co-rotating orbits are different from counter-rotating ones.

In this sense, the description is qualitatively the same, with the main difference being that, for example, in the Kerr metric, $r_{ISCO}$ for the co-rotating/counter-rotating orbits is smaller/larger than the $r_{ISCO}$ of the non-rotating case \cite{hobson2006general, jefremov2015innermost}, and similarly for the IBCO, as shown in Fig. \ref{fig:r_isco_ibco}. Thus, the ranges for $r_{min}$ that we used in the previous section are different for the prograde and retrograde orbits.

Because of this, there are cases in which $r_{min}$ is greater than $r_{ISCO}$ in the co-rotating case but smaller than $r_{IBCO}$ for the counter-rotating one. Consequently, in this situation, while the co-rotating orbits would be in the case where all orbits with arbitrarily large $q$ have periastra arbitrarily close to $r_{min}$, the counter-rotating orbits still include cases where this is not true. In Fig.  \ref{fig:vs_with_rot_co_counter} we compare prograde and retrograde orbits with $r_{min}$ in this range.

\begin{figure}[!h]
\centering
 \includegraphics[width=1\columnwidth]{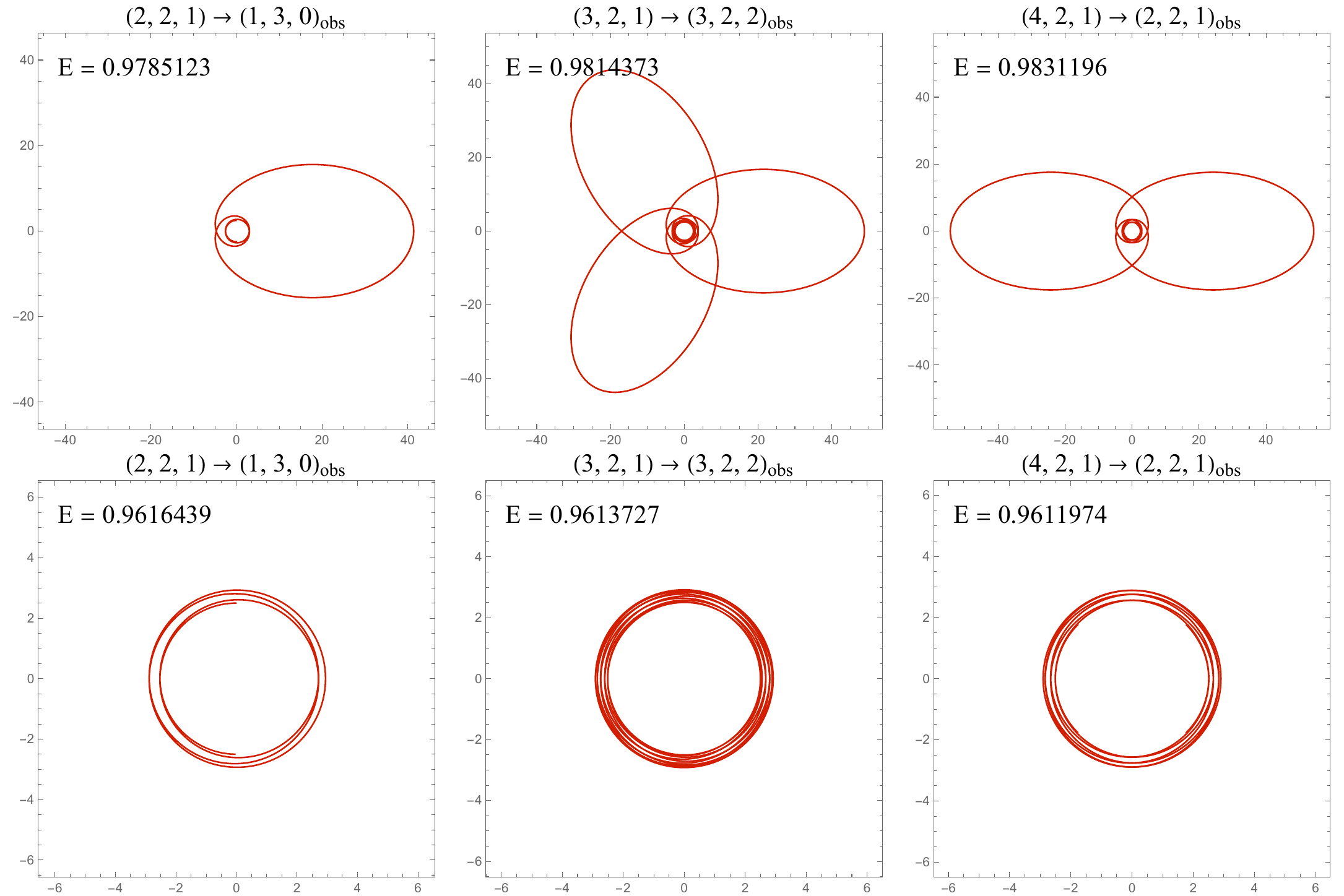}
 \caption{Outer (above) and inner (below) orbits that enter the wormhole described by the rotating Simpson-Visser metric for $r_{min} = 2.5$. All orbits have $a = 0.5$ and $L = 3.2$. The energies are specified in each plot, together with the triplet of integers associated to them and the triplet actually observed $(z,w,v)_{obs}$.}
 \label{fig:vs_with_rotation_wormhole_orbits}
\end{figure}

Moreover, the rotating Simpson-Visser metric describes a two-way traversable wormhole if $r_{min}^2 > (M \pm \sqrt{M^2 - a^2})^2$ and $a < M$ or if $a > 1$ \cite{mazza2021novel}. In this context, the same analysis developed for the non-rotating case can be applied here. So, for instance, it is possible to have two distinct regions, where orbits in the inner region have small values for the eccentricity and are characterized by an associated rational that increases with the energy (or eccentricity), while outer orbits have eccentricities arbitrarily large and exhibit a decreasing associated rational for trajectories with higher energy (or eccentricity). To illustrate this, examples of these two types of orbits are presented in Fig.  \ref{fig:vs_with_rotation_wormhole_orbits}. Of course, co-rotating orbits are different from counter-rotating as exemplified previously, but these traversing orbits follow the same scheme of classification and characteristics as discussed in subsection \ref{subsec:trav_worm}

\section{Conclusion}

In this work, we reviewed a classification scheme proposed by Levin and Perez-Giz \cite{levin2008}, originally applied to closed orbits in the Schwarzschild and Kerr metrics, and extended it to the analysis of geodesics in the Simpson-Visser black bounce model. With that we demonstrated that equatorial closed orbits in both rotating and non-rotating Simpson-Visser metrics can be described by the same topological features as those in the Schwarzschild and Kerr metrics. In this sense, we described the behavior of the trajectories across various ranges of $r_{min}$ and analyzed the effects of an increase of $r_{min}$ in order to compare the trajectories in the Simpson-Visser model and in the Schwarzschild metric (and the rotating case with Kerr), pointing out that in some cases even small increments in $r_{min}$ can lead to significant changes in the orbits.

Understanding the effect that variations in the internal structure of black bounces can have on the orbital motions of particles around them is particularly relevant in dynamical scenarios in which $r_{min}$ can be affected by the accretion of matter. This, in particular, has been observed in numerical evolutions of boson stars that collapse to form nonsingular compact objects \cite{Maso-Ferrando:2023wtz,Maso-Ferrando:2021ngp}. The emergence of a baby universe as a result of the collapse implies the dynamical formation of a black bounce with an evolving $r_{min}$. This $r_{min}$ needs not evolve smoothly from zero to a finite value but, as observed in \cite{Maso-Ferrando:2023wtz,Maso-Ferrando:2021ngp}, it may come into existence with a finite amplitude and then grow and/or decay, potentially affecting the orbital motions of objects around the original object.

Additionally, for cases where the metrics describe a wormhole, we analyzed closed orbits that traverse the throat  and demonstrated that an analogous classification scheme remains valid. Thus, with this geometrical approach and the richness of both rotating and non-rotating Simpson-Visser metrics, we established a natural connection between the description of orbits in different spacetime structures, such as traversable wormholes and regular black holes.

Furthermore, in principle, this framework could be expanded to non-equatorial orbits, relying in an analogy with the generalization for 3D orbits in the Kerr metric that has already been perfomed \cite{levin2009dynamics, grossman2009dynamics}, yielding an elegant description of the dynamics. Also, a general understanding of these orbits is fundamental for observational studies and astrophysical applications. For instance, these aspects are fundamental for modeling extreme mass ratio inspirals (EMRIs) \cite{cardenas2024testing}, and compact objects orbiting supermassive black holes are a plentiful source for gravitational wave astronomy, which may ultimately provide a way to distinguish black holes from alternative exotic compact objects.

\section*{Acknowledgments}
\hspace{0.5cm} ABQ thanks for a PIBIC fellowship (2024–2025) from the Federal University of Ceará. The authors acknowledge financial support from the Spanish Grants   PID2020-116567GB-C21, PID2023-149560NB-C21 funded by MCIU/AEI/10.13039/501100011033 and FEDER, UE, and by CEX2023-001292-S funded by MCIU/AEI.  The paper is based upon work from COST Actions CosmoVerse CA21136 and CaLISTA CA21109 supported by COST (European Cooperation in Science and Technology).



\bibliographystyle{unsrt}
\bibliography{references}

%\begin{thebibliography}{99}


%\end{thebibliography}

%%%%%%%%%%%%%%%%%%%%%%%%%%%%%%%%%%%%%%%%%%%%%%%%%%

%%%%%%%%%%%%%%%%%%%%%%%%%%%%%%%%%%%%%%%%%%%%%%%%%%

% Don't change these lines
% typesetting comment
\label{lastpage}
\end{document}